\documentclass{svjour2}                    
\smartqed  
\usepackage{graphicx}
 \usepackage{mathptmx}      

\usepackage{amsmath}

 \journalname{Journal of Statistical Physics} %

\begin{document}

\title{Commitment versus persuasion in the three-party constrained voter model}

\titlerunning{Commitment versus persuasion in the constrained voter model}   

\author{Mauro Mobilia}


\institute{Mauro Mobilia\at
              Department of Applied Mathematics, School of Mathematics, University of Leeds - Leeds LS2 9JT, U.K.; \\
              Tel.: +44-113-3431591\\
              Fax: +44-113-3435090\\
              \email{M.Mobilia@leeds.ac.uk}           
}

\date{Received: date / Accepted: date}

\maketitle

\begin{abstract}
In the framework of the three-party constrained voter model, where  voters of two radical parties ($A$ and $B$) 
interact  with ``centrists'' ($C$ and $C_{\zeta}$), we study the competition between a persuasive majority and a committed minority.
In this model,  $A$'s and $B$'s  are incompatible  voters that can convince centrists or be swayed by them.
Here, radical voters are more persuasive than centrists, whose sub-population comprises susceptible agents $C$ and a 
fraction $\zeta$ of  centrist zealots $C_{\zeta}$. Whereas $C$'s may
adopt the opinions  $A$ and $B$  with respective rates $1+\delta_A$ and $1+\delta_B$  (with $\delta_A\geq 
\delta_B>0$), $C_{\zeta}$ are committed individuals that always remain centrists. 
Furthermore, $A$ and $B$ voters can become  (susceptible) centrists $C$ with a rate $1$.
The resulting competition between commitment and persuasion is  studied in the mean field limit 
and for a finite population on a complete graph. At mean field level, there is a continuous transition from a 
coexistence phase when $\zeta<\Delta_c=\delta_A/(1+\delta_A)$ to a phase where centrism  prevails  
when $\zeta\geq \Delta_c$. In a finite population  of size $N$,  demographic fluctuations  lead to centrism consensus 
and the dynamics is characterized by the mean consensus time $\tau$. Because of the competition between commitment and persuasion, 
here consensus is reached much slower ($\zeta<\Delta_c$) or faster ($\zeta\geq\Delta_c$) than in the 
absence of zealots (when $\tau \sim N$). In fact, when $\zeta<\Delta_c$ and there is an initial minority of centrists, 
the mean consensus  time  grows  as $\tau\sim  N^{-1/2} e^{N\gamma}$, with $N\gg 1$
and  $\gamma=\delta_A-\zeta(1+\ln{(\delta_A/\zeta)})
+{\cal O}(\delta_A^2)$. The dynamics is thus characterized by a  metastable state where the most persuasive voters and centrists coexist when $\delta_A>\delta_B$, 
whereas all species coexist when  $\delta_A=\delta_B$. When $\zeta\geq\Delta_c$ and the initial density of centrists is low, one finds
 $\tau\sim \ln{N}$ (when $N\gg 1$).
Our analytical findings are corroborated by stochastic simulations.
\end{abstract}

\section{Introduction}
\label{intro}
Opinion dynamics aims at  understanding  how cultural change evolves~\cite{Axelrod,BoundedCompromise,SocRev}.
This issue is closely related to problems arising in various disciplines in the life and behavioral 
sciences~\cite{PopGen},
and the recent years have witnessed a growing activity
 in applying   the tools of statistical physics to their description~\cite{SocRev,voter-variants}.
In models of opinion dynamics,  each 
agent is typically represented by its ``opinion'' state (often labeled as a spin variable~\cite{Liggett}) 
that is updated in response to  the opinion of a local neighborhood,
a mechanism of cultural dynamics documented in sociological studies, see e.g.~\cite{Schelling,Granovetter}.
Basic issues in opinion dynamics concern the time necessary for a consensus to be attained
and the period during which cultural diversity is maintained~\cite{SocRev,voter-variants}.
These questions are usually addressed by considering simple and insightful models, like the 
influential two-state voter model (VM), in which local opinions are copied and spread by imitation, see Sec.~2 below.
It has recently been proposed that the evolution of cultural diversity would be more realistically described by models 
where the seek for {\it consensus} is limited by some form of {\it incompatibility}~\cite{Axelrod,BoundedCompromise}.
This idea has  been investigated in an analytically amenable three-party
constrained voter model (3CVM)~\cite{VKR03,Redner04,MM11} (see also Ref.~\cite{Lanchier12}).

In this work, we generalize the 3CVM to study the competition between  a persuasive majority and a committed 
minority. 
In our model, as in Refs.~\cite{VKR03,Redner04,MM11}, the voters of two radical parties ($A$ and $B$) are incompatible and do 
  not interact among them, but they interact with a 
third species  (``centrists'', $C$ and $C_{\zeta}$). 
Here, radical $A$ and $B$ voters are considered to be more persuasive than centrists,
whose sub-population 
comprises  susceptible agents ($C$) and a small fraction $\zeta$ 
of  ``centrist zealots'' ($C_{\zeta}$). The latter are committed individuals that always remain centrists,
while a susceptible centrist  
adopts the opinion  $A$ with  rate $1+\delta_A$ when it interacts with an $A$-voter,
and the opinion $B$  with  rate $1+\delta_B$ when it interacts with a $B$-voter. 
Furthermore, each $A$ and $B$ individuals can become a (susceptible) centrist with 
rate $1$. The  parameters $\delta_A\geq \delta_B>0$ hence represent the persuasion biases 
toward $A$ and $B$, i.e. the respective persuasion strength of these radical voters.
The dynamics is thus characterized by a
competition between the persuasion strength of $A$ and $B$
voters opposed by the resistance of centrist zealots ($C_{\zeta}$).
The questions that we address with this model are the following:
{\it (i) When and how does a committed minority prevail against a persuasive 
majority? (ii) How does the mean consensus time vary with the fraction of  zealots 
and the persuasion biases?}

These questions are answered by studying the generalized 3CVM's properties  in the mean field limit and for a 
finite population of size $N$ 
on a complete graph. 
In the absence of fluctuations, we find that more than one opinion coexists 
when $\zeta$ is below a critical 
threshold $\Delta_c=\delta_A/(1+\delta_A)$, while centrism prevails when $\zeta\geq\Delta_c$.
The dynamics is markedly different when the size of the population is finite and the
demographic fluctuations drive the system into centrism consensus in a time that 
depends non-trivially on $\delta_A, \delta_B, \zeta$ and  $N$.
The mean consensus time 
in a large population with an  initial minority of centrists is here shown 
to grow 
dramatically  as 
$\tau\sim  N^{-1/2} e^{N\gamma(\delta_A,\zeta)}$ when  $\zeta<\Delta_c$, and  as $\tau\sim \ln{N}$ when  $\zeta\geq \Delta_c$.
These results, derived using the Fokker-Planck equation and the WKB approach, are in stark contrast with the mean consensus time 
 obtained in the absence of   centrist zealots (when $\tau\sim N$~\cite{Redner04,MM11}). Our findings, 
corroborated by stochastic simulations, thus demonstrate that the presence of centrism zealotry either significantly slows down
 ($\zeta<\Delta_c$) or speeds up ($\zeta\geq \Delta_c$) the approach to consensus.

While some aspects of the influence of committed agents in opinion dynamics have been considered in the literature, e.g. in Refs.~\cite{zealots1,zealots2,zealots3,GalamJacobs,committed} 
(see Sec.~2), this work differs from earlier studies in various respects: (i) We here consider a {\it three-party (four-state)} model
whereas most of previous works, like Refs.~\cite{zealots1,zealots2,zealots3,GalamJacobs,committed}, focused on two-state systems.
(ii) Here, we study the competition between the degree of commitment in the population {\it and} the  persuasion 
strength of two types of voters (three independent parameters), while 
no persuasion bias was considered in \cite{zealots3,GalamJacobs,committed}.
(iii) Furthermore, whereas  most of the previous results were obtained
in the mean field limit~\cite{GalamJacobs} or  by means of numerical  simulations~\cite{committed}, 
 the exponent $\gamma(\delta_A,\zeta)$ is here obtained {\it analytically}.

This paper is organized as follows: 
Next Section is dedicated to a brief review of the voter model and some of its variants, while the
3CVM with centrist zealots is introduced in Section 3. The mean 
field dynamics of such a model is discussed in Section 4.  The section 5 is dedicated to the stochastic formulation of the dynamics 
in a finite population. The mean consensus time in the case of identical persuasion biases is studied in Section 6. 
In particular,
the case of the long-lived coexistence of opinions and metastability 
is studied in terms of the Fokker-Planck equation (Sec.~6.1.1) and with the WKB method (Sec.~6.1.2). 
The mean consensus time in the general case of asymmetric persuasion biases is analyzed in Section 7.
In the final section, our findings are summarized and our conclusions presented.
\section{A brief review of the voter model and some of its variants: zealotry, commitment and incompatibility}
\label{sec0}
As a contribution to a special issue on the {\it ``applications of statistical mechanics to social phenomena''}, this
 work is dedicated to the study of the competition between commitment and  persuasion in the three-party constrained voter 
model. Before presenting our new findings, it is useful to begin by providing some background material in the form of a brief review of some basic models and results.

Since the pioneering works by Schelling and Granovetter~\cite{Schelling,Granovetter}, the importance of relying on individual-based models
to relate ``micro-level  to macro-level interactions''~\cite{Granovetter} in social dynamics has been recognized.
As an inspiring example of this type of modeling approach,
Schelling showed how {\it homophily}, that is the tendency of an individual to being bond with neighbors sharing similar
characteristics, is  important  to understand the formation of social segregation~\cite{Schelling}.
\\
In this context,  the voter model (VM)~\cite{Liggett} is one of the simplest and most paradigmatic {\it individual-based} models
of opinion dynamics. The VM describes how a socially interacting population of
individuals possessing a discrete set of states (``opinions'') 
evolves toward consensus as the result of 
interactions between neighbors and local fluctuations of the population composition.
The VM is closely related to the  Ising model of statistical physics with the Glauber kinetics at zero 
temperature~\cite{Glauber,KRB}\footnote{These models are {\it equivalent} on one-dimensional lattices, 
but not in higher dimensions where the 
Ising-Glauber model follows a {\it majority} rule, whereas the VM evolves according to a {\it a proportional rule}. 
As a consequence, while there is a surface tension in the two-dimensional Ising-Glauber model this is not the case for the VM,
see~\cite{SocRev,KRB} and references therein.}, and also to the Moran model~\cite{Moran} commonly
used to describe evolutionary dynamics in the life and behavioral sciences~\cite{PopGen}\footnote{The voter and Moran models are
equivalent on regular graphs but lead to markedly different dynamics on degree-heterogeneous networks, see e.g.~\cite{VMnets}.}.
In the classic two-state VM, each node of a graph is occupied by a ``voter'' in one of the two possible state, e.g.
 denoted $A$ or $B$  (or $+/-$, $\uparrow/\downarrow$, or ``leftist/rightist'').
 These voters can be interpreted as having no self confidence
and changing their state by simply adopting the opinion of a neighbor.
 Hence, in its simplest form the VM evolves according to the following rules:
\begin{itemize}
 \item Pick a random voter.
\item The selected voter adopts the opinion of one of its random neighbor.
\item These steps are repeated until a consensus (where all individuals are of the same opinion, either all leftists or all rightists) is necessarily reached.
\end{itemize}
In the language of statistical physics, these rules and the VM 
define a Markov chain with absorbing states that is aptly described by a master equation~\cite{Gardiner}.
One of the most appealing features of the VM is its mathematical tractability. 
Indeed, it is one of the rare  models of non-equilibrium statistical physics to be exactly solvable in any dimensions 
on regular lattices, and many of its properties are known analytically, see e.g.~\cite{SocRev,KRB,Gardiner}. The most significant features of the VM are (i) the probability to attain a specific consensus state, (ii) the mean 
time to reach a consensus, and (iii) the two-point correlation function ${\cal G}(r,t)$. 
\begin{enumerate}
 \item[i.]  The consensus probability $E_A(x)$ (also called ``exit probability'', or ``fixation probability'' by analogy with the evolutionary dynamics literature~\cite{PopGen}) is the probability 
that a finite system  consisting of an initial density $x$ of $A$-voters
eventually reaches the consensus state  where all voters are in the state $A$. 
As the average opinion, also called ``magnetization'',  is conserved by the VM on any degree-regular graphs (like lattices 
and complete graphs), and since the VM always reaches consensus, one 
has $E_A(x)=x$ on any regular graphs.
\item[ii.] On degree-regular graphs, it is known that the mean consensus time $\tau$ (also called ``mean exit time'' 
or ``mean fixation time'') in a population of $N$ voters scales as $\tau\sim N^2$ in one dimension ($d=1$) 
and as $\tau\sim N\ln{N}$ 
in two dimensions ($d=2$),
whereas in dimensions $d>2$  one has $\tau \sim N$ (complete graphs correspond to $d=\infty$). 
This implies that $d=2$ is the upper critical dimension of the VM.
Moreover, it has recently been shown that
 the mean consensus time scales sublinearly with $N$  on degree-heterogeneous graphs characterized by nodes of high-degree, 
i.e. $\tau\sim N^{\alpha}$,  where
$0<\alpha<1$ depends non-trivially on the network's degree-distribution~\cite{VMnets}~\footnote{It has also been shown that 
 a class of models with voter-like dynamics and describing cooperation dilemmas  are characterized by 
anomalous metastability on scale-free networks, with the mean consensus time and exit probability exhibiting a stretched exponential dependence on the 
the population size and an exponent depending non-trivially of the degree distribution~\cite{AM2012}.}.
\item[iii.] Significant insight into the spatial distribution of opinions is provided by the two-point correlation
function ${\cal G}(r,t)$. This quantity informs on the probability that two voters separated by a distance $r$ are in
 the same state at time $t$.
On  lattices of dimension $d$, ${\cal G}(r,t\to \infty)$ decays asymptotically as $r^{d-2}$ in high dimensions ($d>2$).
In low dimensions the spatial organization of the VM is characterized by {\it coarsening}, i.e. the slow formation of 
growing domains of a single opinion. As a consequence, the two-point correlation
function asymptotically approaches the value $1$ in low dimensions. More precisely, when $t\to \infty$ and $0<r<\sqrt{t}$, one finds
$1-{\cal G}(r,t)\sim r/\sqrt{t}$ in $d=1$ and $1-{\cal G}(r,t)\sim \ln{r}/\ln{t}$ in $d=2$~\cite{Coarsening}.
\end{enumerate}
While the VM has proved to be insightful and influential, it relies on oversimplified assumptions like the total lack of self-confidence 
of all voters and the unavoidable formation of a consensus. In fact, since  Granovetter's seminal work on ``threshold models''~\cite{Granovetter}, 
the influence on social dynamics of the population's {\it heterogeneous} response to stimuli  is well established.
In the opinion dynamics context, as a step toward the investigation of a simple model  describing the dynamics of a 
heterogeneous population with different levels of confidence, we have proposed to investigate the 
two-state voter model in the 
presence of ``zealots''~\cite{zealots1,zealots2,zealots3}. The zealotry is implemented by assuming that a small fraction of the 
population consists of {\it committed} individuals (called zealots) that either favor a  specific state ($A$ or $B$) as 
in Refs.~\cite{zealots1,zealots2}, or adopt an opinion that never changes as in Ref.~\cite{zealots3}.
As a consequence of zealotry, the magnetization 
 is not conserved in the VM with zealots, and the main features of these models are the following:
\begin{itemize}
 \item  In Ref.~\cite{zealots1} it was shown that on  low-dimensional lattices 
(when $d\leq 2$) a single zealot with a favored opinion (i.e. a voter with a bias toward one opinion, 
but whose state is {\it not} fixed) imposes the consensus in its favored state to an infinite group of voters,
 whereas in higher dimension the magnetization is non-uniform and decays with the distance from the zealot.
Furthermore, the unanimity state imposed by the zealot is approached algebraically in one dimension, as $t^{-1/2}$, and much slower
 in two 
dimensions ($\sim 1/\ln{t}$). It has also been shown that on a finite lattice of size $N=L^d$ (where $L$ is the lattice linear size), the time 
necessary for the zealot's opinion to spread is $N^2$ is $d=1$, $N\ln{N}$ in $d=2$, and $N$ in higher dimensions.
\item Ref.~\cite{zealots2} was dedicated to the study of the 
 VM in the presence of a small group of $n$ zealots, each with a favored opinion and a specific bias toward 
its preferred state. The competition between zealots of different types was shown to prevent the formation of consensus and 
to lead to a non-trivial fluctuating steady state whose properties were investigated by exploiting a formal analogy 
with
the electrostatic potential generated
by $n$ classical point charges. The approach toward the steady state was shown to be algebraic ($\sim t^{-1/2}$) and logarithmic 
($\sim 1/\ln{t}$) in one and two dimensions, respectively, and to be characterized by the formation of growing  domains
when the density of zealots is very small, i.e. $n/N\ll 1$. In this case, the size of the single-opinion domains was found to grow
 with the system size but to never span entirely the system.
\item The voter model with a finite (but small) fraction $n/N$ of zealots that never change opinion has been studied in
Ref.~\cite{zealots3}. The study was carried out on complete graphs and on low-dimensional lattices with  randomly distributed
 zealots (of both types). 
It was thus shown that in all cases a small fraction of zealots is effective in maintaining a reactive state
 characterized by a Gaussian magnetization distribution with a width that decays as $n^{-1/2}$.
\end{itemize}
The influence of committed agents in opinion dynamics has also been considered in other two-opinion dynamics models, see e.g. 
Refs.~\cite{GalamJacobs,committed,Commit,Heterogeneity}: 
\begin{itemize}
 \item  The case of a committed minority of ``inflexibles'' in a 
two-state majority-rule model has been considered at mean field level 
in Ref.~\cite{GalamJacobs},  where it was shown that an equal densities of inflexibles of each type
prevent consensus from being achieved.
\item The authors of Ref.~\cite{committed} 
studied how the mean consensus  time varies with the fraction  of committed individuals
in a two-opinion variant of the naming game~\cite{NG} called the {\it binary agreement model}.
In the model of Ref.~\cite{committed}, the population initially consists of a majority  of individuals of one opinion and 
a  fraction $p$ of individuals of the other species. As in Ref.~\cite{zealots3},  the latter are committed individuals that never change state
 and  impose their consensus after a mean time showed to grow exponentially
 or logarithmically with the population size, depending on whether
 $p$ is below or above a critical threshold~\cite{committed}.
\item The authors of \cite{Commit} considered 
the {\it confident voter model}, in which each voter has
two levels of commitment (``confident'' and ``unsure'') for each of the two possible opinions.
The confident VM has been studied in the mean field limit, where the mean consensus time is $\tau \sim \ln{N}$, whereas 
in one and two dimensions the mean consensus time scales respectively as $\tau \sim N^2$ and $\tau \sim N^{3/2}$~\cite{Commit}.
\item Furthermore, the {\it heterogeneous voter model} in which each agent has its own intrinsic rate to change state has been 
considered in Ref.~\cite{Heterogeneity}, where it was found  that
the time until consensus is reached is much longer in the heterogeneous VM than in the classic voter model.
\end{itemize}
It is also worth noting that 
the influence on cooperation dilemmas of zealot-like  individuals  has recently been studied in the framework of evolutionary games~\cite{EGT-heterogeneous}.

\vspace{0.15cm}

In addition to voter-like models with committed individuals, there are various prototypical
opinion dynamics models, whose outcome is characterized by a lack of consensus. Influential models of this type are the 
multiple-state Axelrod model~\cite{Axelrod} and the bounded compromise model~\cite{BoundedCompromise}.
The key feature preventing consensus in these models is a form of {\it incompatibility}: when the opinions of two agents are 
too different, they are deemed to be  incompatible and there is no interaction between such individuals. 
This can lead to cultural fragmentation and a frozen stationary state 
where a mixture of incompatible states coexist. In its essence, the feature of incompatibility is captured by the so-called 
three-party  constrained voter 
model (3CVM) that can be regarded as a discrete three-state version of the bounded compromise
model~\cite{VKR03,Redner04}. In the 3CVM, there is no interaction between species $A$ (``leftists'') and
$B$ (``rightists'') that are are incompatible. However, $A$- and $B$-voters
interact with a third species $C$, and thus compete to impose their own
consensus according to the following rules:
\begin{itemize}
 \item A  centrist $C$ can become an $A$-voter or a $B$-voter with rate $1+\delta'$ (and $0\leq |\delta'|\leq 1$);
 \item  $A$ and $B$ voters can become  centrists  $C$ with a rate $1$;
\item   $A$ and $B$ voters do not interact: $AB \to AB$,
\end{itemize}
where $\delta'$ is a bias favoring either $A$ and $B$ (when $\delta'>0$), or $C$ (when $\delta'<0$).
The 3CVM admits three absorbing fixed points, corresponding to the consensus with $A, B$ and $C$, 
and a ``polarization line'' along which a frozen mixture of non-interacting $A$'s and $B$'s coexist.
This 3CVM  has been solved  analytically on a complete graph and
the probabilities of reaching each consensus state and the polarization
line  were determined, both when  $\delta'=0$~\cite{Redner04} and  in the case
$\delta'\neq 0$~\cite{MM11}. Within an approach based 
on the Fokker-Planck equation~\cite{Gardiner}, it was also shown that the mean exit times in the
 3CVM scale linearly with the population size 
$N$ when the intensity of the bias is weak (i.e. when $N|\delta'|\ll 1$ and $N\gg 1$)~\cite{Redner04,MM11}.
The 3CVM has also investigated on regular lattices and found to exhibit
 slow  kinetics in low dimensions~\cite{VKR03} (see also \cite{Lanchier12}).

Aiming to study how the maintenance of cultural diversity is affected by the competition between persuasion and commitment, 
in this work we consider a three-party (four-state) opinion dynamics model that combines the main features of both the 
3CVM and VM with zealots. The resulting model, whose detailed specification is given in the next section, is here used to study the competition between two 
 radical opinions ($A$ and $B$) that are incompatible and the voters of a third party that are less persuasive but on average 
more committed than $A$'s and $B$'s.

\section{The three-party constrained voter model with centrist zealots}
\label{sec1}

In this work, we introduce and study a three-party constrained voter model with centrist zealots on a complete graph 
(used for its tractability and because it  is arguably a natural first choice in the context of social dynamics). Like the 3CVM~\cite{VKR03,Redner04,MM11}, this model is characterized by a  population of $N$ individuals,
 $j$ are of species $A$, $k$ of type $B$ and $\ell$ are ``susceptible'' centrists (species $C$). 
Moreover, the population also comprises $\ell_{\zeta}$ committed centrists (zealots), denoted $C_{\zeta}$,  that never 
{\it change opinion}, as in \cite{zealots3}. Hence, one has
$N=j+k+\ell+\ell_{\zeta}$, with a fixed fraction $\zeta \equiv \ell_{\zeta}/N$ of zealots $C_{\zeta}$. 
In the language of the voter model (see Sec.~2), the individuals of species $A$ (``leftists'') and $B$ (``rightists'')
are voters of two radical parties,  while supporters of the third party (``centrism'') 
are either  susceptible centrists (species $C$)
or centrist zealots (type $C_{\zeta}$).
As in the 3CVM, radical opinions  are incompatible and there are no interactions between 
$A$'s and $B$'s, but 
these voters interact with a centrist neighbor according to the following evolutionary rules:
\begin{enumerate}
 \item A {\it susceptible centrist} $C$ can become an $A$-voter with rate $1+\delta_A$ and a 
$B$-voter with rate $1+\delta_B$;
 \item {\it $A$- and $B$-voters} can become susceptible centrists with a rate $1$;
\item  {\it $A$- and $B$-voters} do not interact: $AB \to AB$;
 \item {\it Centrist zealots} $C_{\zeta}$ always remain in this state.
\end{enumerate}
In the spirit of the VM and 3CVM, the dynamics of the system is thus implemented as follows:
\begin{itemize}
 \item[(i)] At each time-step a pair of (``neighboring'')
voters is randomly picked;
\item[(ii)] if an   
$A/B$-centrist pair is picked, the population composition changes
according to the following schematic moves:
\begin{eqnarray}
\label{rules}
  A  C &\to A  A \quad &{\rm with \ rate} \ 1+\delta_A   \nonumber \\
  B  C &\to B  B \quad &{\rm with \ rate} \ 1+\delta_B    \\
  A  C &\to C  C \quad &{\rm and} \quad   A  C_{\zeta} \to C  C_{\zeta} \quad {\rm with \ rate} \ 1 \; \nonumber \\
  B  C &\to C  C  \quad &{\rm and} \quad   B  C_{\zeta} \to C  C_{\zeta} \quad {\rm with \ rate} \ 1; \;  \nonumber
\end{eqnarray}
\item[(iii)]  if the randomly picked pair of neighbors consists of two radical voters ($AA, BB$ or $AB$) or two centrists 
($C C, C  C_{\zeta}$ or $C_{\zeta} C_{\zeta}$ ), the composition of the population does not change;
\item[(iv)] these steps are iterated until centrism consensus is necessarily reached (see below).
\end{itemize}

In this work, we assume the existence of {\it persuasion biases} (or persuasion strengths) 
$\delta_A>0$ and  $\delta_B > 0$ toward opinions $A$ and 
$B$, respectively. The quantities $\delta_{A/B}$ measure the biases toward opinions $A/B$ (against centrism)
and thus reflect that $A$ and $B$ voters are
 more persuasive than 
centrists. 
Here, without loss of generality, we assume that
$\delta_A\geq \delta_B >0$.
The limiting case $\ell_\zeta= \delta_A=\delta_B=0$
corresponds to the 3CVM without zealots and no bias that was studied in Ref.~\cite{Redner04},
while the 
3CVM with a symmetric bias but without zealots  of Ref.~\cite{MM11}
is recovered when $\ell_\zeta=0, \delta_A=\delta_B\neq 0$.

From the rules (\ref{rules}), we expect a subtle competition
between $A$ and $B$ voters, favored by their persuasion biases and seeking to impose their consensus, 
and centrist zealots that resist the spread of $A$'s and $B$'s  and  strive to 
 promote centrism. We are particularly interested in such a competition when 
centrists are initially
in the minority and the population is finite. {\it In this situation, we will study the circumstances
under which a committed minority can prevail against a persuasive majority.
Another question of great interest concerns how
the competition between persuasion and commitment leads to 
a long-lived (metastable) coexistence state along with the maintenance of a form of cultural diversity.}
These issues are studied in the mean field limit (Sec.~4) and  in  finite populations (Secs.~5-7).

\section{Mean field analysis}
\label{sec1.2}
%
%
\begin{figure}
\begin{center}
\includegraphics[width=4.65in, height=2.15in,clip=]{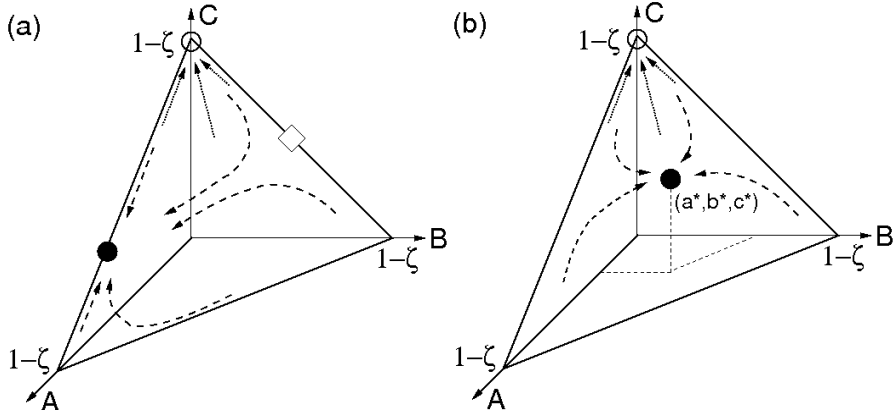}
\caption{Schematic phase portrait of the mean field dynamics (\ref{RE}). (a)  Asymmetric biases $\delta_A>\delta_B>0$:
when $\zeta<\Delta_c$, only the coexistence state ($\bullet$) corresponding to a population of $A$-voters and centrists
 is stable (flows in dashed arrows), while  the fixed point associated with centrism consensus ($\circ$) is unstable. 
When  $\zeta<\delta_B/(1+\delta_B)$ (as in this sketch), there is also an unstable coexistence state ($\diamond$) consisting of
$B$-voters and centrists. When  $\zeta>\Delta_c$, the only stable fixed point corresponds to 
the centrism consensus  (dotted arrows).
(b) Identical biases $\delta_A=\delta_B=\delta>0$: the fixed point $(a^*, b^*, c^*)$ ($\bullet$) associated with the coexistence
of the three parties is stable  when $\zeta<\Delta_c$ (flows in dashed arrows) and unstable otherwise 
(dotted arrows). 
See text.
}
  \label{MFdiag1}
\end{center}
\end{figure} 

At mean field level, one assumes a population of infinite size ($N\to \infty$) and ignore any 
random fluctuations. Hence,
the densities $a\equiv j/N, 
b\equiv k/N$ and $c\equiv \ell/N$, and the fraction $\zeta \equiv 
\ell_{\zeta}/N$ of (centrist) zealots, are treated as continuous variables obeying the following rate equations (REs):
\begin{eqnarray}
 \label{RE}
\frac{d}{dt}a(t)&=& a(t)\,[\delta_A \,c(t)-\zeta] \nonumber\\
\frac{d}{dt}b(t)&=& b(t)\,[\delta_B \,c(t)-\zeta].
\end{eqnarray}
The total population size being conserved, one has $c(t)=1-\zeta-(a(t)+b(t))$. Substituting this expression into
(\ref{RE}) yields two coupled equations whose properties are discussed below, first in the case of 
asymmetric biases and then when the biases $\delta_A$ and $\delta_B$ are identical. In fact, 
we readily notice from (\ref{RE}) that in the absence of bias ($\delta_A=\delta_B=0$),
the system quickly evolves toward a population comprising only centrists, whereas an interesting situation arises when
$\delta_A \geq \delta_B>0$.
\subsection{Mean field dynamics with asymmetric biases}
When $\delta_A>\delta_B>0$, the mean field equations (\ref{RE}) are characterized by three fixed points $(a,b,c)\stackrel{t \to \infty} 
 {\longrightarrow}  (a_i^*,b_i^*,c_i^*)$ with $i\in (1,2,3)$:
\begin{eqnarray}
 \label{fixed1}
 (a_1^*,b_1^*,c_1^*)&=& (0,0,1-\zeta),\nonumber\\
 (a_2^*,b_2^*,c_2^*)&=& (1-\zeta-\frac{\zeta}{\delta_A},0,\frac{\zeta}{\delta_A}), \\
 (a_3^*,b_3^*,c_3^*)&=& (0,1-\zeta-\frac{\zeta}{\delta_B},\frac{\zeta}{\delta_B}).\nonumber
\end{eqnarray}
The  non-trivial fixed point  $(a_2^*,b_2^*,c_2^*)$ physically exists and  is asymptotically 
stable when  $\zeta<  \Delta_c\equiv \delta_A/(1+\delta_A)$. Otherwise, 
$(a_1^*,b_1^*,c_1^*)$ is the only stable fixed point
when $\zeta\geq   \Delta_c$. Furthermore, there is also
another coexistence fixed point $(a_3^*,b_3^*,c_3^*)$  when $\zeta< \delta_B/(1+\delta_B)$, 
but a stability analysis reveals that it is always unstable (saddle point). 
As illustrated in Fig.~\ref{MFdiag1}(a),  the mean field 
dynamics (\ref{RE}) therefore predicts that the system evolves toward the  {\it stable fixed point} 
\begin{eqnarray}
 \label{REc}
(a(t),b(t),c(t)) \stackrel{t \to \infty}  {\longrightarrow} (a^*,b^*,c^*)
\equiv
\begin{cases}
   (a_2^*,b_2^*,c_2^*)& \text{when} \quad  \zeta< \Delta_c \; \text{and}  \;   \delta_A>\delta_B\\
(a_1^*,b_1^*,c_1^*)     & \text{when} \quad 
 \zeta\geq \Delta_c.
  \end{cases}
\end{eqnarray}
This means that in the case of  asymmetric persuasion biases with $\delta_A>\delta_B$,  
when $\zeta$ is below the critical  threshold  $\Delta_c=\delta_A/(1+\delta_A)$,
the mean field dynamics leads to a coexistence state where the most persuasive voters ($A$'s) coexist
with centrists. In this case, $B$-voters are absent form the final state.
On the other hand, when $\zeta\geq \Delta_c$ the mean field dynamics
evolves toward centrism consensus, with a population consisting only of susceptible and committed centrists.  
\subsection{Mean field dynamics with a identical bias $\delta_A=\delta_B=\delta>0$}
\label{sec1.2}
When the persuasion bias toward opinions $A$ and $B$ is identical, i.e. $\delta_A=\delta_B=\delta>0$,
the rate equations (\ref{RE}) can be solved exactly. In fact, by adding the equations of (\ref{RE})
and using $a(t)+b(t)=1-\zeta-c(t)$, we obtain 
\begin{eqnarray}
 \label{REc}
\frac{d}{dt}c(t)=-\frac{d}{dt}(a(t)+b(t))=-\delta (1-\zeta -c(t))[c(t)-\zeta /\delta].
\end{eqnarray}
This equation can be readily solved and by denoting the initial densities 
$(a(0),b(0),c(0))=(x,y,z)$, when $\zeta\neq \Delta_c$, one finds:
\begin{eqnarray}
 \label{REc}
c(t)
= \frac{1-\zeta +K\zeta \, {\rm exp}\left(\delta (1-\zeta -\zeta /\delta)t\right)
}{1+K \, {\rm exp}\left(\delta (1-\zeta -\zeta /\delta)t\right)},
\end{eqnarray}
where  $K\equiv \frac{1-z-\zeta }{\delta z -\zeta }$. 
When the biases toward $A$ and $B$ are identical, one has 
 $\frac{d}{dt}\left(\frac{a(t)}{b(t)}\right)=0$,  which implies that 
$a(t)/b(t)=x/y$ is  conserved by (\ref{RE}). Using this property together
with (\ref{REc}), we obtain
$a(t)=\left(\frac{x}{x+y}\right)\left[1-\zeta -c(t)\right]$ and $b(t)=(y/x)a(t)$.
According to (\ref{REc}),
the mean field dynamics is therefore characterized by an exponentially quick approach to the 
steady state  densities $(a^*,b^*,c^*)$, where
\begin{eqnarray}
 \label{cinf}
c(t)\stackrel{t \to \infty}  {\longrightarrow} c^*=
\begin{cases}
   \zeta/\delta & \text{if} \quad  \zeta< \Delta_c \; 
\\
   1-\zeta       & \text{if} \quad  \zeta> \Delta_c,
  \end{cases}
\end{eqnarray}
and 
\begin{eqnarray}
 \label{abinf}
a(t)=\frac{x}{y}b(t)\stackrel{t \to \infty}  {\longrightarrow} a^*=\frac{x}{y}  b^*=
\begin{cases}
   \frac{x(1-\zeta(1+\delta^{-1}))}{K(\delta z -\zeta)} & \text{if} \quad  \zeta< \Delta_c  \;\\
   0      & \text{if} \quad  \zeta> \Delta_c,
  \end{cases}
\end{eqnarray}
as sketched in Fig.~\ref{MFdiag1}(b).
This means that when the persuasion biases are identical, and if the density $\zeta$ of centrist zealots is 
below $\Delta_c$, the mean field analysis predicts the stable coexistence of $A$ and $B$ voters along with 
centrists. If $\zeta>\Delta_c$, the final state is again composed  only of (susceptible and committed) centrists. 
Furthermore, in the critical case $\zeta= \Delta_c$, one has 
the long-time behavior $c(t)-(1+\delta)^{-1}\sim (\delta t)^{-1}$ when $t\to \infty$. In this critical case
one is left with a  density $\zeta/\delta=(1+\delta)^{-1}$ of susceptible centrists along with a fraction
$\zeta$
of  zealot centrists, but no  $A$ and $B$ voters.

According to these mean field results, summarized in Fig.~\ref{MFdiag1}, 
when the fraction $\zeta$ of zealots in the population is low,
the dynamics generally reaches a stationary state where more than one species coexists. Hence, the mean field analysis predicts the transition 
from a coexistence
and ``multicultural''
 phase ($\zeta<\Delta_c$) to a phase  dominated by centrism ($\zeta\geq\Delta_c$)
at the critical value $\zeta=\Delta_c$.
As discussed in the next sections, this picture
changes drastically when the population is of finite size and 
 demographic fluctuations need to be taken into account.

\section{Stochastic dynamics in a finite population}
\label{sec2.1}
When the population is of finite size $N$, demographic (random) fluctuations significantly alter the mean field dynamics and 
 ultimately impose the 
centrist consensus. The latter is associated with the system's unique absorbing state that consists of $N-\ell$ susceptible centrists ($C$)
and $\ell_\zeta$ centrist zealots ($C_{\zeta}$).
While the system's fate is known, we are here interested in the dependence of the consensus (exit) time
  on the
 population size and on the parameters $\delta_A$, $\delta_B$ and 
$\zeta$. In particular, we study the circumstances under which  
cultural diversity is maintained (long-lived coexistence), and when  consensus is quickly reached.
These questions are addressed by modeling the evolution with continuous-time birth-death processes.

In the general case of  different persuasion biases ($\delta_A\geq \delta_B$), 
at each interaction   the number of $A$ and $B$ voters can increase or decrease by one 
(i.e. $j \to j \pm 1$ and $k \to k \pm 1$) with rates $T_{j,k}^{A\pm}$ and $T_{j,k}^{B\pm}$, respectively (see (\ref{rules})).
The population's composition therefore evolves according to a {\it bivariate}
birth-death process  described by the probability 
${\cal P}_{j,k}(t)$ that at time $t$ there are $j$ voters of type $A$ and a number $k$ of $B$ voters in the population. 
 This probability obeys the  master equation~\cite{Gardiner}:
\begin{eqnarray}
 \label{MEbiv}
\frac{d{\cal P}_{j,k}(t)}{dt}&=&
T_{j-1,k}^{A+}{\cal P}_{j-1,k}(t)+T_{j+1,k}^{A-}{\cal P}_{j+1,k}(t)+
T_{j,k-1}^{B+}{\cal P}_{j,k-1}(t)\nonumber\\
&+&T_{j,k+1}^{B-}{\cal P}_{j,k+1}(t)-
[T_{j,k}^{A+}+T_{j,k}^{A-}+T_{j,k}^{B+}+T_{j,k}^{B-}]{\cal P}_{j,k}(t),
\end{eqnarray}
where the state space is bounded since $j,k \in [0,N-\ell_\zeta]$ and, to account for the fact that $j=k=0$ corresponds
to the absorbing state with only centrists ($C$'s and $C_{\zeta}$'s), one has $T_{0,0}^{\pm A}=T_{0,0}^{\pm B}=0$ and
${\cal P}_{j,k}(t)=0$ for $j>N-\ell_\zeta$ and $j<0$, and for $k>N-\ell_\zeta$ and $k<0$. According to (\ref{rules}), the 
transition rate $T_{j,k}^{A+}$ is 
associated with the reaction of the pair $AC\to AA$
and $T_{j,k}^{B+}$ with $BC\to BB$, while   $T_{j,k}^{A-}$ and $T_{j,k}^{B-}$ are 
respectively associated with $AC/C_{\zeta}\to C C/C_{\zeta}$ and $BC/C_{\zeta}\to C C/C_{\zeta}$.
Since the probabilities of picking the  pairs $AC$ and $AC_{\zeta}$
are respectively $j\ell/N(N-1)$  and $j\ell_{\zeta}/N(N-1)$, and similarly for the pairs $BC$ and $BC_{\zeta}$, the transition rates
read
\begin{eqnarray}
 \label{TRbiv}
T_{j,k}^{A+}&=& (1+\delta_A) \frac{j(N-\ell_{\zeta}-j-k)}{N(N-1)} \quad , \quad
T_{j,k}^{A-}= 
 \frac{j(N-j-k)}{N(N-1)},
\nonumber\\
T_{j,k}^{B+}&=& (1+\delta_B) \frac{k (N-\ell_{\zeta}-j-k)}{N(N-1)} \quad , \quad
T_{j,k}^{B-}= 
 \frac{k(N-j-k)}{N(N-1)},
\end{eqnarray}
where we have used $j+k=N-\ell_{\zeta}-\ell$.

The mathematical treatment greatly simplifies in the case of identical 
persuasion biases, when $\delta_A=\delta_B=\delta>0$.
In this case,  centrists $C$ interact in the same manner with $A$ and $B$ voters (see (\ref{rules})) and 
at each time step their number can change by $\pm 1$ with transition rates $T_{\ell}^{\pm}$, respectively. The 
dynamics can thus be mapped onto a {\it single-variate} process described
in terms of the probability $P_{\ell}(t)$ 
of finding $\ell$ susceptible centrists in the population 
at time $t$. This quantity obeys the master equation
\begin{eqnarray}
 \label{ME}
\frac{dP_{\ell}(t)}{dt}=T_{\ell-1}^{+}P_{\ell-1}(t)+T_{\ell+1}^{+}P_{\ell+1}(t)-\left[T_{\ell}^{+} + T_{\ell}^{-}\right]
P_{\ell}(t),
\end{eqnarray}
where, to account for the absorbing boundary conditions, we impose
$T_{N-\ell_\zeta}^{\pm}=0$ and
$P_{\ell}(t)=0$ for $\ell>N-\ell_\zeta$ and $\ell<0$. 
According to (\ref{rules}), and from the above discussion, the interaction $XC \to XX$, with $X\in \{A,B\}$,
is associated with the rate $T_{\ell}^{-}=T_{j,k}^{A+}+T_{j,k}^{B+}$.
In the same manner, $T_{\ell}^{+}$
corresponds to the reactions $XY \to CY$, with $Y\in \{C,C_{\zeta}\}$, and thus $T_{\ell}^{+}=T_{j,k}^{A-}+T_{j,k}^{B-}$. 
The transition rates of the master equation  (\ref{ME}) therefore read
\begin{eqnarray}
 \label{TR}
T_\ell^{+}=
 \frac{(N-\ell_{\zeta}-\ell)(\ell+\ell_{\zeta})}{N(N-1)}, \quad \text{and} \quad
T_\ell^{-}= 
(1+\delta)\frac{(N-\ell_{\zeta}-\ell)\ell}{N(N-1)}.
\end{eqnarray}

Here we are chiefly interested in the mean consensus  time (MCT)  $\tau$ necessary 
 to reach the absorbing boundary $(j,k,\ell)=(0,0,N-\ell_{\zeta})$ starting from a population with initial densities 
$(a(0), b(0),  c(0))=(x, y, z)$, 
while $\zeta$ remains fixed. The MCT will be studied both for  (\ref{MEbiv},\ref{TRbiv}) and 
(\ref{ME},\ref{TR}). 
In the case of identical persuasion biases, the MCT of the birth-death process (\ref{ME},\ref{TR}) 
will be obtained analytically. The direct treatment of the process (\ref{MEbiv},\ref{TRbiv})
describing the 3CVM with asymmetric persuasion biases is difficult, but its dynamics is exactly reproduced by the 
Gillespie algorithm used in our simulations~\cite{Gillespie}. Furthermore, in Sec.~7 we will 
use the analytical results obtained in Sec.~6 for 
the unbiased case to analytically characterize the MCT also when $\delta_A>\delta_B$.
\section{Mean consensus time in the 3CVM with centrist zealots and identical persuasion biases}
In this section, we study the mean consensus time (MCT) of the birth-death process defined by (\ref{ME},\ref{TR}) 
when the persuasion biases are identical, i.e. $\delta_A=\delta_B\equiv\delta>0$, and the  population size $N\gg 1$ is large (but finite). 
We consider the continuum limit and treat 
 $z=\ell/N$ and $\zeta=\ell_{\zeta}/N$  as continuous variables. The
 transition rates (\ref{TR}) therefore become
\begin{eqnarray}
 \label{TRcont}
T^{+}(z)\equiv
(1-\zeta-z)(z+\zeta), \quad \text{and} \quad
T^{-}(z)\equiv 
(1+\delta) (1-\zeta-z)z.
\end{eqnarray}
In this setting, the MCT obeys the backward master equation~\cite{Gardiner,MM11}
\begin{eqnarray}
 \label{backME}
\left(
T^{+}(z)+T^{-}(z)
\right)\tau(z)= \Delta +
T^{-}(z)\tau(z-\Delta) + T^{+}(z)\tau(z+\Delta),
\end{eqnarray}
where the time has been rescaled according to $t\to t/N$ 
and  $\Delta\equiv N^{-1}$
(the time increment 
 $\Delta$ hence matches how  the density of $C's$ changes at each interaction).

In principle (\ref{backME}) is solvable, but the formal exact solution is an unwieldy and non-enlightening expression~\cite{Gardiner}.
A more tractable and insightful approach is provided by system size expansions of (\ref{backME}) in powers of $\Delta$. 
In the realm of the diffusion theory~\cite{Kimura,Gardiner}, one Taylor expands (\ref{backME}) to second order in $\Delta$ and
obtains the 
backward Fokker-Planck equation (bFPE)
\begin{eqnarray*}
 (T^{+}(z) -T^{-}(z)) \tau'(z)
 + ((T^{+}(z) + T^{-}(z))/2N) 
\tau''(z)=-1
\end{eqnarray*}
 This bFPE  holds for all single-coordinate birth-death processes~\cite{Gardiner}
and, with (\ref{TRcont}), here explicitly reads
\begin{eqnarray}
\label{bFPE}
(1-\zeta -z)\left[-(\delta z - \zeta)\,\frac{d\tau(z)}{dz}
+\frac{1}{2N}\left\{(2+\delta)z +\zeta\right\}\frac{d^2\tau(z)}{dz^2}
\right]=-1.
\end{eqnarray}
This equation is supplemented by the absorbing and reflecting boundary conditions $\tau(1-\zeta)=0$ and $\tau'(0)=0$, respectively. 
The condition $\tau(1-\zeta)=0$ accounts for the fact that the dynamics ends when 
$z=1-\zeta$ (i.e. $\ell=N-\ell_{\zeta}$) and the centrism consensus (absorbing state) is reached. The reflecting condition 
$\tau'(0)=0$
ensures that the system cannot be ``exited'' by the end $z=0$ (i.e. $\ell=0$)~\cite{Gardiner}.
Eq.~(\ref{bFPE}) gives the MCT in the framework of the diffusion theory, which is certainly an accurate 
approximation  when the intensity of the deterministic drift is much less than the intensity of the fluctuations 
($\propto N^{-1/2}$)~\cite{Kimura,Assaf2010,AM2011}.
The formal solution of (\ref{bFPE}) can be obtained by standard methods and reads~\cite{Gardiner}:
\begin{eqnarray}
\label{sol1}
\tau(z)=2N \, \int_{z}^{1-\zeta} dv \; e^{N{\cal F}(v)}
\int_{0}^{v} du \, \frac{e^{-N{\cal F}(u)}}{\left( 1-\zeta-u\right)\left(\zeta+(2+\delta)u\right)},
\end{eqnarray}
where
\begin{eqnarray}
\label{F}
{\cal F}(x)&\equiv& nx-m\ln{\left[(2+\delta)x + \zeta\right]}, \\
\text{with} \quad
n &\equiv& \frac{2\delta}{2+\delta} \quad \text{and} \quad   
m\equiv \frac{4\zeta(1+\delta)}{(2+\delta)^2}. \nonumber
\end{eqnarray}
The result (\ref{sol1}), is valid in both regimes $\zeta<\Delta_c$ and 
 $\zeta\geq\Delta_c$, and is particularly relevant 
when the persuasion bias $\delta$ and the fraction  $\zeta$ of centrists zealots are both ``small'' 
(e.g. in the range ${\cal O}(N^{-1})-{\cal O}(N^{-1/2})$), which leads to an effective competition between
the deterministic nonlinear drift and the diffusive noise. 
It is worth emphasizing that in the case of identical biases,
centrists do not discriminate between
$A$ and $B$ voters and  the MCT is thus a function of  $x+y=1-\zeta-z$ (total initial density of radical voters) or, 
 equivalently, of  the initial density $z$ of $C$'s. A particularly interesting situation arises when
$z\to 0$ and the system consists of a vanishing minority of
centrists and an overwhelming majority of radical ($A$ and $B$) voters.
Below, we show that two very different types 
of behaviors arise: a form of cultural diversity is maintained over a long period of time
 when $\zeta<\Delta_c$, whereas centrism consensus is quickly reached when  $\zeta\geq\Delta_c$.
\subsection{Long-lived coexistence and consensus time}
\begin{figure}
\begin{center}
\includegraphics[width=2.19in, height=1.69in,clip=]{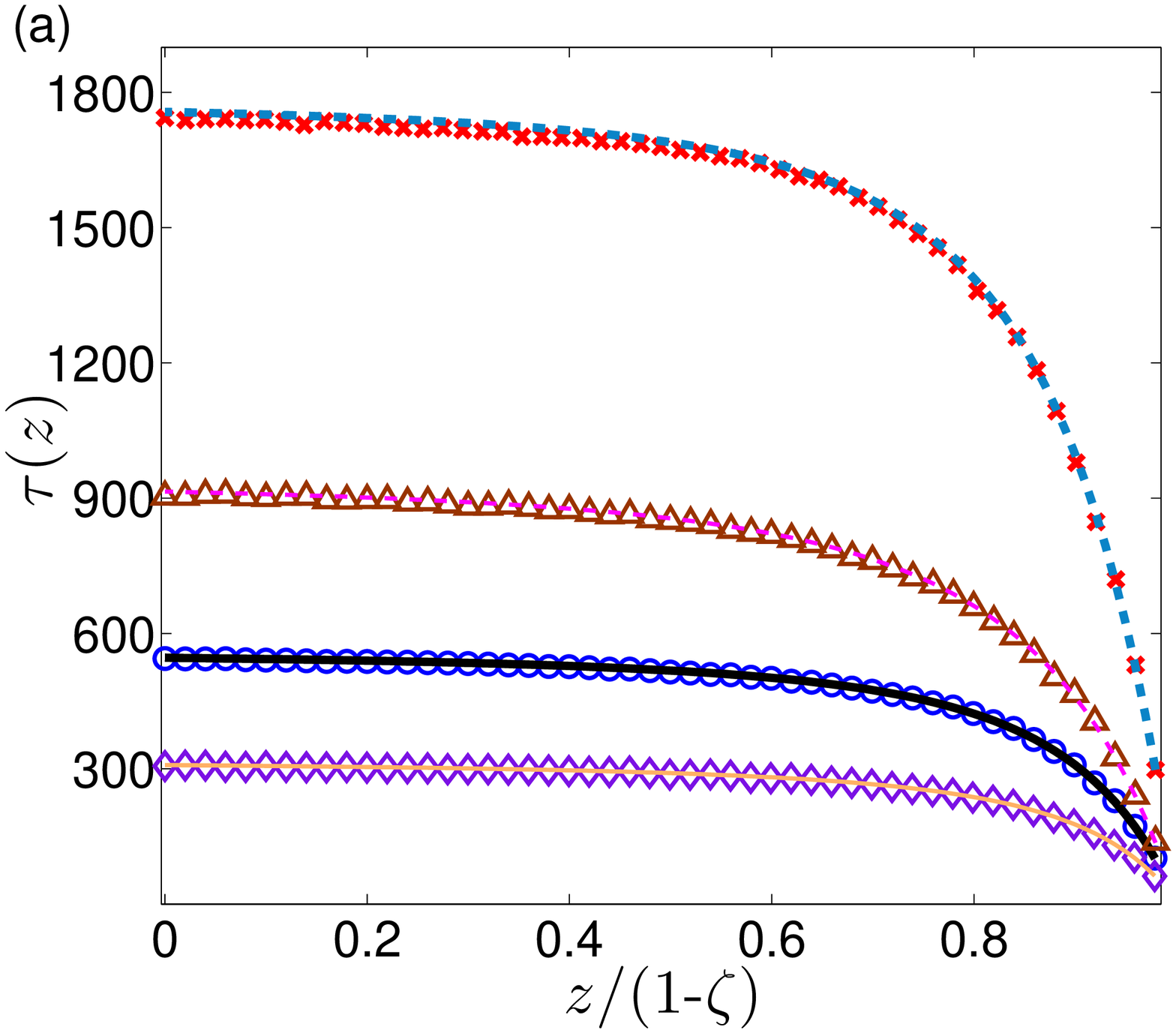}
\includegraphics[width=2.20in, height=1.7in,clip=]{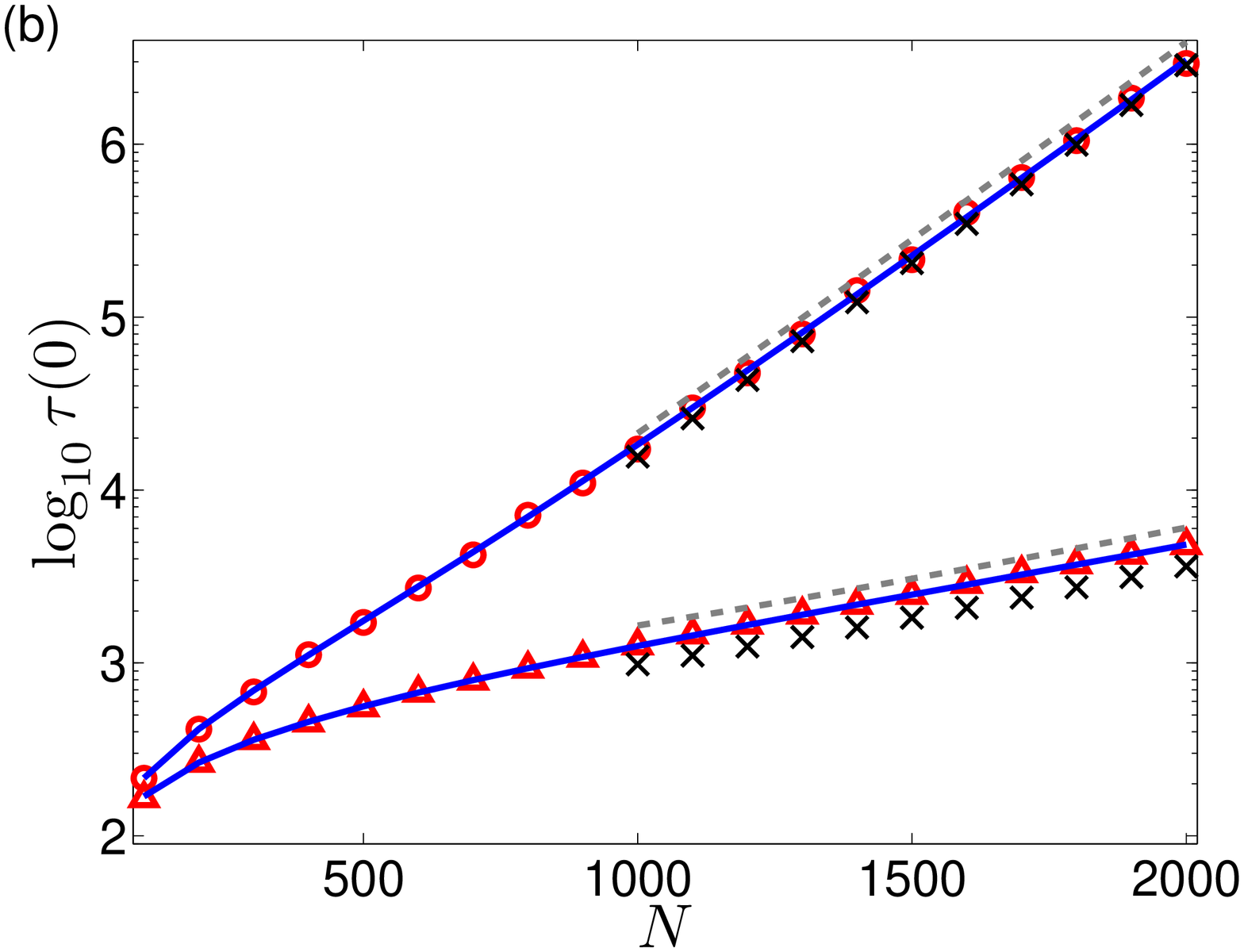}
\caption{{\it (Color online)}. Functional dependence of the MCT  when $\zeta<\Delta_c$ and $\delta_A=\delta_B=\delta$.
(a) $\tau(z)$ as function of the (rescaled) initial concentration of susceptible centrists $z/(1-\zeta)$. Results of stochastic simulations (symbols) 
averaged over 
$5\times 10^4$ samples
are compared with formula (\ref{sol1}) (curves)
for $(\delta,\zeta)=(0.04,0.02)$ ($\times$, thick dashed), $(0.035,0.02)$ ($\triangle$, thin dashed),
$(0.06,0.04)$ ($\circ$, thick solid), and $(0.08,0.06)$ ($\diamond$, thin solid).
The population size is $N=500$.  
(b) Logarithm of the MCT as  function of  $N$ 
with $\zeta=0.02$ and $\delta=0.04 (\circ)$ and
 $\delta=0.03 (\triangle)$
 starting from a system without susceptible centrists: $\log_{10}{\tau(0)}$
has been computed using stochastic simulations (symbols, averaged over 1000 samples with $N=100 - 2000$), and is 
compared with the predictions of (\ref{sol1}) (solid)
and the WKB results (\ref{MET2}) ($\times$).  The (rescaled) asymptotic predictions (\ref{sol4}) are shown as dashed lines.
}
  \label{Fig2}
\end{center}
\end{figure}

We first study the situation where
 $\zeta<\Delta_c$ or, equivalently, 
$\delta>\zeta/(1-\zeta)$. 
In this case,
 the mean field rate equations (\ref{RE}) predict the coexistence of 
the three opinions with a density
$\zeta/\delta$ of susceptible centrists (see Fig.~\ref{MFdiag1}(b)). Here, by paying special attention to the interesting situation where $\zeta <\delta \ll 1$ and
$N \gg 1$, we analyze how this picture is altered by random fluctuations and how
centrism consensus is ultimately reached. From the results  of stochastic simulations reported in Fig.~\ref{Fig2}.
it appears that the MCT decreases monotonically with the density $\zeta$ of centrist zealots  and increases with the persuasion bias 
$\delta$. The MCT also appears to grow
steeply (almost exponentially) with the population size $N$, see Fig.~\ref{Fig2}(b).
Moreover, the results of Fig.~\ref{Fig2}(a) indicate that $\tau \simeq \tau(0)$
when the initial number of susceptible centrists is small ($z\ll 1-\zeta $), which
implies that in such a regime
the MCT is essentially independent of the initial density $z$. 
Below,  these observations are substantiated by the analysis of the functional dependence of the MCT on 
 $\delta,\zeta$ and $N$,
first by using the Fokker-Planck equation and then with the WKB method.
\subsubsection{Mean consensus time with the Fokker-Planck equation when  $\zeta<\Delta_c$}
In this section, we analyze the predictions of (\ref{sol1}) when $\zeta<\Delta_c$ and $z\ll 1$,
 and obtain the leading contribution of the  mean consensus time 
in the framework of the Fokker-Planck equation (\ref{bFPE})
 using a saddle-point approximation.

\vspace{0.2cm}

\begin{figure}
\begin{center}
\includegraphics[width=3.0in, height=2.0in,clip=]{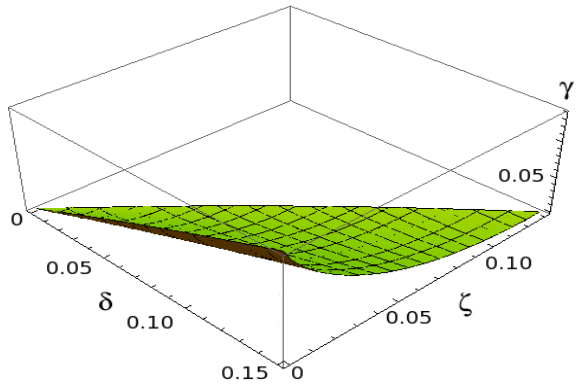}
\caption{{\it (Color online)}. Functional dependence of  $\gamma$ on the parameters $\delta$ and $\zeta$
when $\zeta<\Delta_c$ according to (\ref{sol3}): $\gamma$ grows and decreases monotonically with 
$\delta$ and $\zeta$, respectively.
}
  \label{gammafig}
\end{center}
\end{figure}

When $N\gg 1$ and $\zeta<\Delta_c$,  
the function ${\rm exp}{(-N{\cal F}(u))}$ has an isolated peak at $u^*\equiv\zeta/\delta$ and one can Taylor-expand  ${\cal F}(u)$ around $u^*$ which yields
${\rm exp}{(-N{\cal F}(u))}={\rm exp}{(-N\left[{\cal F}(u^*)+(u-u^*)^2 {\cal F}''(u^*)/2\right]}$.
We can use this expansion to   evaluate  the inner integral of the right-hand-side (RHS) of (\ref{sol1}) with a saddle-point 
approximation,  yielding
\begin{equation*}
\int_{0}^{v} du \, \frac{e^{-N{\cal F}(u)}}{\left( 1-\zeta-u\right)\left(\zeta+(2+\delta)u\right)}
\simeq 
e^{-N{\cal F}(\zeta/\delta)} \, \int_{-\infty}^{\infty} du \; e^{-(N/2)(u-\zeta/\delta)^2{\cal F}''(\zeta/\delta)}.
\end{equation*}
In setting the boundaries  to $\pm \infty$ we have assumed that the peak is sufficiently separated from the 
absorbing boundary, i.e. we henceforth assume that $\zeta/\delta$ is well separated from the values $0$
and $1$. By performing the Gaussian integral, the expression (\ref{sol1}) becomes
\begin{eqnarray}
\label{sol2}
\tau(z)&\simeq&
\sqrt{\frac{2\pi N}{\zeta(1+\delta)}}\, \frac{e^{-N{\cal F}(\zeta/\delta)}}{1-\zeta(1+\delta^{-1})}\, \int_z^{1-\zeta} dv\, 
e^{N{\cal F}(v)}.
\end{eqnarray}
As  ${\cal F}(v)$ is an increasing function of 
$v$
on $\zeta/\delta<v\leq 1-\zeta$,  the main contribution to the integral on the RHS of (\ref{sol2})
arises from $v\simeq 1-\zeta$ and therefore $\int_z^{1-\zeta} dv\, e^{N{\cal F}(v)}\simeq
\int_z^{1-\zeta} dv\, e^{N[{\cal F}(1-\zeta)-(1-\zeta-v){\cal F}'(1-\zeta)]}\sim e^{N{\cal F}(1-\zeta)}/N$. Hence, with (\ref{sol2}), 
and provided that the initial population does not mainly consist of
susceptible centrists (i.e. if $z\ll 1$),  
the leading contribution to the MCT in the realm of the diffusion approximation is 
\begin{eqnarray}
\label{sol3}
\tau &\sim& N^{-1/2}\, e^{N\gamma(\delta,\zeta)} \equiv
N^{-1/2}\, e^{N\left({\cal F}(1-\zeta)-{\cal F}(\zeta/\delta)\right)},
\end{eqnarray}
where the functional dependence of the exponent 
$\gamma(\delta,\zeta)\equiv {\cal F}(1-\zeta)-{\cal F}(\zeta/\delta)$ is illustrated in Fig.~\ref{gammafig}. 

As we are particularly interested in the limit of weak persuasion biases and small fraction of zealots, i.e.
 $\zeta<\delta\ll 1$ (with $z\ll 1-\zeta$), the expansion of the exponent 
$\gamma$ yields $\gamma(\delta,\zeta)= 
\delta- \zeta(1 +\ln{\left(\delta/\zeta\right)})
+{\cal O}(\delta^3)$. With (\ref{sol3}), one therefore obtains the following asymptotic result when $N\gg 1$: 
\begin{eqnarray}
\label{sol4}
\tau&\sim&
N^{-1/2}\, e^{N\left(\delta- \zeta \right)}\, \left(\frac{\zeta}{\delta}\right)^{N\zeta}.
\end{eqnarray}
The concise result (\ref{sol4}) provides the leading contribution to the MCT in the realm of the diffusion theory 
and is valid as long as $\zeta<\delta\ll N^{-1/3}$.
According to this result, the leading contribution to the MCT grows exponentially with the population size and with the difference
$\delta-\zeta$ (in a manner that is independent of the initial condition $z$, see below), and decreases with $\zeta$.
The exponential growth  of the MCT with $N(\delta-\zeta)$ is confirmed by the stochastic simulations reported in Fig.~\ref{Fig2}, 
where we find an excellent agreement with the predictions of (\ref{sol1}). The results of Fig.~\ref{Fig2}(a) also show 
that the MCT depends ``weakly'' on the initial density $z$ of susceptible centrists: in fact, 
the MCT is found to significantly deviate from the value 
$\tau(0)$ only when $z/(1-\zeta)> 0.7$, i.e. in the special case where there is already  an  overwhelming initial majority of centrists.
In Figure \ref{Fig2}(b), we see that the asymptotic behavior of the MCT when $N(\delta-\zeta)\gg 1$ is aptly captured by (\ref{sol4}).

Furthermore, to understand the influence of the term $(\zeta/\delta)^{N\zeta}$ on the RHS of (\ref{sol4}),
 it is worth considering the transformation 
$(\delta,\zeta) \to (\delta',\zeta')=(\delta+\alpha,\zeta+\alpha)$, where $\alpha$ is a given (small) real number.
Under such a transformation, the exponential term of  (\ref{sol4}) is left unaltered, but $(\zeta/\delta)^{N\zeta}\to ((\zeta+\alpha)
/(\delta+\alpha))^{N(\zeta+\alpha)}$
and therefore  the MCT decreases if $\alpha>0$ and increases otherwise. This nontrivial interplay 
between $\delta$ and $\zeta$ is confirmed by the results of 
Fig.~\ref{Fig2}(a) and is illustrated in Fig.~\ref{gammafig} where we see that  the exponent
$\gamma$ decays along the line $\delta-\zeta=$constant.
\subsubsection{Metastability and WKB treatment when $N(\delta-\zeta) \gg 1$ }
The asymptotic behavior  (\ref{sol4}) of the MCT has been derived from the solution of the Fokker-Planck equation (\ref{bFPE}),
by assuming  that $\zeta< \delta \ll 1$.
In this section, we use a complementary approach in terms of the Wentzel-Kramers-Brillouin (WKB) theory~\cite{WKB,Dykman} to obtain the 
MCT to the  next-to-leading order in the limit where $N(\delta-\zeta) \gg 1$  when $\zeta<\Delta_c$.

While the mean field analysis (\ref{RE}) 
 predicts the stable coexistence of all opinions in this regime, when the population 
is large but finite, the system lingers around the coexistence state (with a density $\zeta/\delta$ of $C$'s) 
before centrism consensus is eventually reached.
The dynamics is   thus characterized by
{\it metastability}, a phenomenon that is well described by  the WKB method~\cite{WKB}.
The WKB approximation  is an asymptotic theory frequently used in
the semi-classical treatment of quantum mechanics~\cite{WKB} that is also useful to study stochastic 
processes~\cite{Dykman}. Recently, there have been numerous applications of the WKB approach to study problems of metastability arising in 
 population dynamics, as well as in evolutionary games and population genetics, see e.g.~\cite{Assaf2010,Escudero,AM2011}.
As the WKB method has never been used in the framework of opinion dynamics to the best of our knowledge, we  here follow 
Refs.~\cite{Assaf2010,AM2011} and outline the essence of the method and show how it can be used to compute the MCT when
 $N(\delta-\zeta) \gg 1$.

The WKB treatment is based on a size expansion of  the master equation (\ref{ME}) using an 
exponential ansatz for the probability distribution (see (\ref{ansatz}) and Ref.~\cite{WKB}).
It relies on two main assumptions: (i) the population size is large (i.e.
$N\gg 1$) and one can work in the continuum limit;
 (ii) the system quickly relaxes toward the metastable state 
and, from there, reaches the 
absorbing (consensus) state after a time growing dramatically with $N$~\cite{Dykman,Escudero,Assaf2010,AM2011}.
Here, this implicitly means that we assume that the initial density of centrists is low enough 
($z\ll 1-\zeta$),
to ensure that the metastable state is always reached prior to consensus (see Fig.~\ref{MFdiag1}).
In this setting, the mean consensus time $\tau_{\rm WKB}$ is  the 
decay time of the metastable state described by the centrism
 {\it quasi-stationary probability distribution} (QSD) $\pi_{\ell}$. The QSD is 
obtained by approximating 
$P_{0\leq \ell\leq N-\ell_{\zeta}}\simeq \pi_{\ell} e^{-t/\tau_{\rm WKB}}$ and $P_{N-\ell_{\zeta}}\simeq 
1-e^{-t/\tau_{\rm WKB}}$~\cite{Assaf2010,AM2011}.
Using these expressions in the master equation (\ref{ME}), one obtains
the quasi-stationary master equation
\begin{equation}
\label{QSD}
T^+_{\ell-1}\pi_{\ell-1}+T^-_{\ell+1}\pi_{\ell+1}-\left[T^+_{\ell}+T^-_{\ell}\right]\pi_{\ell}=0,
\end{equation}
where 
an exponentially small term $\pi_{\ell}/\tau_{\rm WKB}$ has been neglected
and we have used the transition rates (\ref{TR}).
The decay time of the metastable state coincides with the MCT and is given by the flux of probability into the absorbing 
state, i.e.
\begin{equation}
\label{tau1}
\tau^{-1}_{\rm WKB}= T^{+}_{N-\ell_{\zeta}-1}\pi_{N-\ell_{\zeta}}\simeq \left|\frac{dT^{+}(1-\zeta)}{dz}\right|\frac{\pi(1-\zeta)}{N},
\end{equation}
where  we have used the continuum limit $z\equiv \ell/N$, $\pi_\ell\equiv \pi_{Nz}=\pi(z)$ 
and $T^{+}_{N-\ell_{\zeta}-1}\simeq |dT^{+}(1-\zeta)/dz|/N$, with the continuous rates (\ref{TRcont})~\cite{Assaf2010,AM2011}.
According to (\ref{tau1}), the calculation of $\tau$ requires to solve the equation (\ref{QSD}) for the QSD. This can fruitfully be done  
using the WKB approach~\cite{Dykman}, which is based on the ansatz
\begin{equation}
\label{ansatz}
\pi(z)\simeq {\cal A} e^{- N S(z)-S_1(z)},
\end{equation}
where $S(z)$ and $S_1(z)$ are respectively the system's ``action'' and ``amplitude'', 
while ${\cal A}$ is a normalization constant. The 
action is found by substituting (\ref{ansatz}) into (\ref{QSD})  
and by keeping the leading order in $N$, which
yields $T^{+}(z)[e^{S'(z)}-1]+T^{-}(z)[e^{-S'(z)}-1]=0$, with $S'\equiv dS/dz$~\cite{Dykman,Escudero,Assaf2010,AM2011}. 
With (\ref{TRcont}),  one thus finds 
\begin{equation}
\label{action}
S(u)=-\int^u\ln [T^{+}(\nu)/T^-(\nu)]\,d\nu=
u\, \ln{\left(\frac{(1+\delta)u}{u+\zeta}\right)}
-\zeta\ln{(u + \zeta)}.
\end{equation}
\\
 The constant ${\cal A}$ is determined by normalization of
 the Gaussian approximation
$\pi(z)\sim {\cal A} e^{- N S(z^*)-(N/2)S''(z^*)(z-z^*)^2}$
around the metastable state $z^*=\zeta/\delta$. 
A Gaussian integration yields ${\cal A}\sim  e^{N S(\zeta/\delta)}$ and, 
to leading order, $\pi(z)\sim e^{-N(S(z)-S(\zeta/\delta))}$.
With (\ref{tau1},\ref{action}), the leading contribution to the MCT therefore reads~\cite{Escudero,Assaf2010,AM2011}:
\begin{eqnarray}
 \label{MET1}
\tau_{\rm WKB} \sim e^{N\gamma_{\rm WKB}(\delta,\zeta)}\equiv e^{N(S(1-\zeta)-S(\zeta/\delta))}=
\left[(1+\delta)(1-\zeta) \, \left(\frac{\zeta}{\delta(1-\zeta)}\right)^{\zeta}\right]^N
\end{eqnarray}
It is worth noting that, to low orders in $\delta$ and $\zeta$,  the exponent $\gamma_{\rm WKB}$ of (\ref{MET1}) reads
\begin{eqnarray*}
\gamma_{\rm WKB}(\delta,\zeta)\equiv
 S(1-\zeta)-S(\zeta/\delta)&=&(1-\zeta)\ln{(1+\delta)(1-\zeta)}+\zeta\ln{\zeta(1+\delta^{-1})}
\nonumber\\ &=&
\delta-\zeta(1+\ln{(\delta/\zeta)})-(\delta^2-\zeta^2)/2 + {\cal O}(\delta^3).
\end{eqnarray*}
Hence,  the exponents $\gamma$  of the Fokker-Planck result (\ref{sol3})  and  $\gamma_{\rm WKB}$ 
of the WKB treatment ({\ref{MET1}}) coincide to first (leading) order in $\zeta<\delta\ll 1$,
but their  next-to-leading-order contributions are different.

The calculation of the subleading correction to the MCT (\ref{MET1}) is feasible but more involved.
In fact, it necessitates $S_1(z)$ and requires to match the ansatz (\ref{ansatz}) with the solution of the (\ref{QSD}) linearized
about the absorbing boundary $z=1-\zeta$. Here, we quote the final result
and refer to 
Ref.~\cite{Assaf2010} for a detailed calculation and a related discussion~\footnote{One slight difference with the generic result 
given in Ref.~\cite{Assaf2010} lies in the fact that here the absorbing boundary is at $z=1-\zeta$. Furthermore, 
in the final result (\ref{MET2})
 we have chosen the same timescale as in Sec.~6.1.1 and divided (\ref{tau1}) by $N$. 
This allows a direct comparison with the results of the Fokker-Planck treatment
and differs by a factor $N^{-1}$ from that of \cite{Assaf2010,AM2011}.}:
\begin{equation}
\label{MET2}
\tau_{\rm WKB} \simeq\sqrt{\frac{2\pi}{N}\left(\frac{1-\zeta}{\zeta}\right)} 
\frac{\delta \left[(1+\delta)(1-\zeta)\right]^N}{(\delta(1-\zeta)-\zeta)^2} \, \left(\frac{\zeta}{\delta(1-\zeta)}\right)^{N\zeta}.
\end{equation}
As illustrated by Fig.~\ref{Fig2}(b), when the population size is sufficiently large  and $N(\delta-\zeta)\gg 1$,
(\ref{MET2}) is in good agreement with the predictions of the FPE and 
with the results of stochastic simulations.
Two additional remarks are in order:
\begin{enumerate}
 \item[i.] In contrast to the prediction (\ref{sol1}) of the diffusion approximation,
 the 
results (\ref{MET1},\ref{MET2}) for $\tau_{\rm WKB}$ only depend on the parameters $\delta$ and $\zeta$ but are 
{\it independent} of the initial density of centrists $z$, and  thus are to be compared with $\tau(0)$. 
This is because the MCT is here computed {\it from} the metastable state, which is certainly always legitimate when the initial population 
does not consist of an overwhelming majority of centrists (i.e. provided that $z,\zeta\ll 1$).
\item[ii.] The accuracy of the WKB approximation improves when $N(S(1-\zeta)-S(\zeta/\delta)) \gg 1$, 
see e.g.~\cite{Assaf2010,AM2011},
as illustrated in Fig.~\ref{Fig2}(b) where it is shown that the WKB results are more accurate 
when $\delta=0.04$ (and $N>1000$)  than for $\delta=0.03$.
\end{enumerate}

\subsection{Centrism consensus time when  $\zeta\geq \Delta_c$}
\begin{figure}
\begin{center}
\includegraphics[width=2.25in, height=1.75in,clip=]{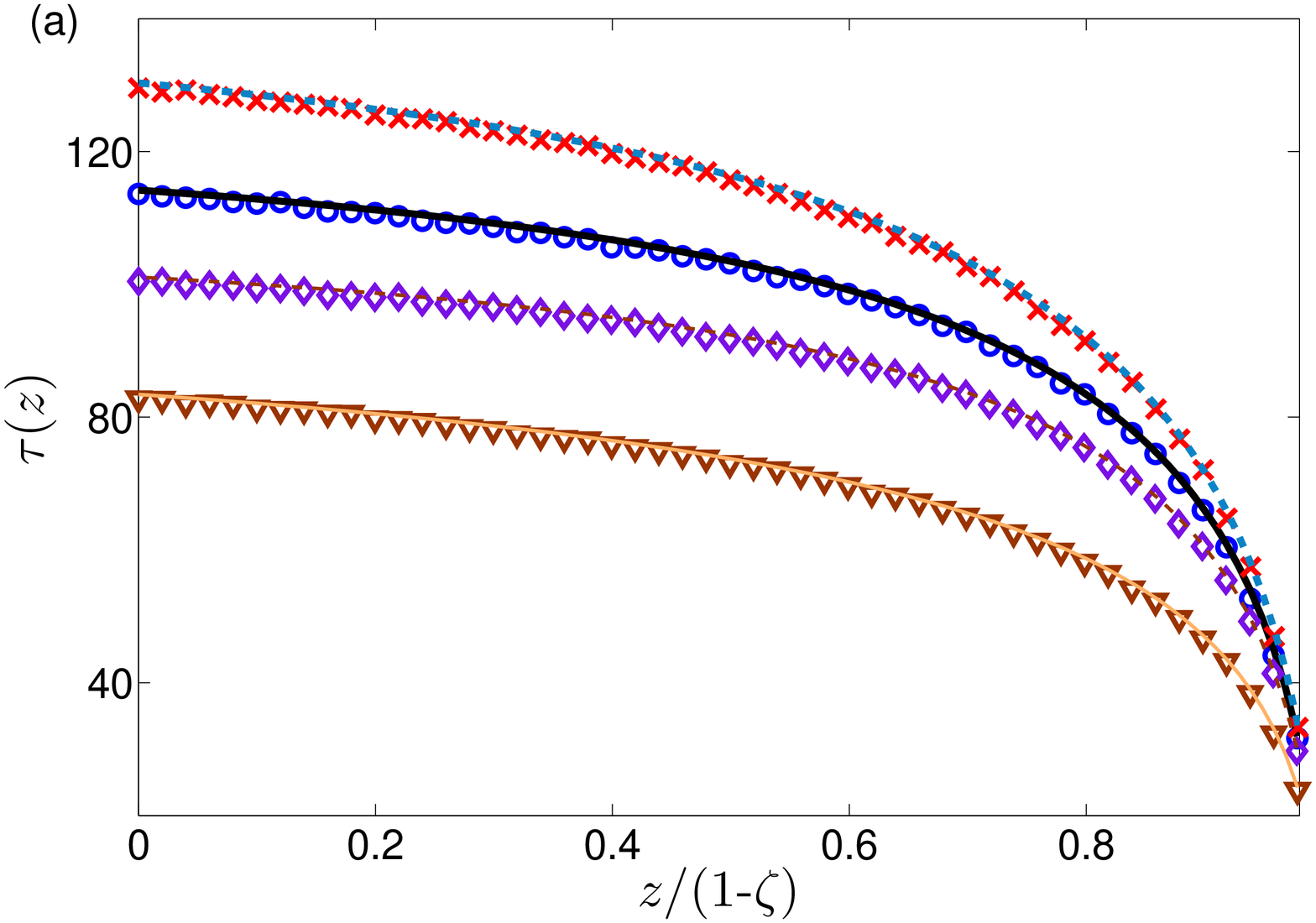}
\includegraphics[width=2.25in, height=1.75in,clip=]{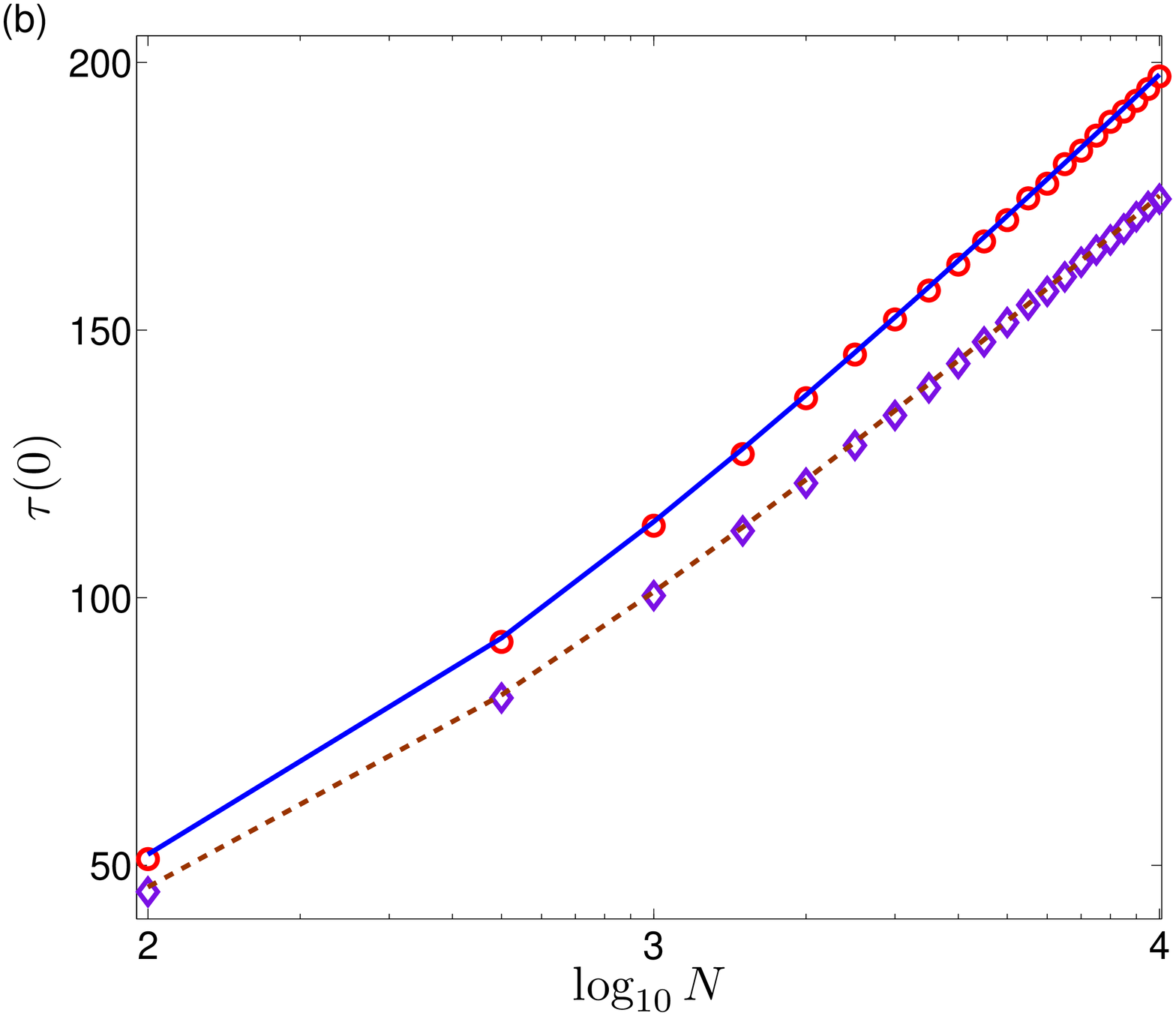}
\caption{{\it (Color online)}.  Functional dependence of the MCT when $\zeta\geq\Delta_c$ and $\delta_A=\delta_B=\delta$.
(a) $\tau(z)$ as function of the (rescaled) initial concentration of susceptible centrists $z/(1-\zeta)$
for various values of $\delta$ and $\zeta$. The results of stochastic simulations (symbols), averaged over $5\times 10^4$ samples
are compared with formula (\ref{sol1}) (curves) for $(\delta,\zeta)=(0.04,0.06)$($\times$, thick dashed), $(0.06,0.08)$($\circ$, thick solid),
$(0.08,0.10)$($\diamond$, thin dashed) and $(0.035,0.08)$ ($\nabla$, thin solid).
(b)  The MCT as function of  $\log_{\rm 10}N$, with $N=100-10000$:
 starting from a system without susceptible centrists with $(\delta,\zeta)=(0.06,0.08)$ ($\circ$) and $(\delta,\zeta)=(0.08,0.10)$ ($\diamond$),  $\tau(0)$
has been computed using stochastic simulations (symbols) and the formula (\ref{sol1}) (solid/dashed lines).
}
  \label{Fig4}
\end{center}
\end{figure} 
When $\zeta\geq \Delta_c$, the rate equation (\ref{REc}) is not  characterized by an interior 
fixed point and the mean field rate equations (\ref{RE}) predict 
that centrism consensus is quickly reached (after a short  coexistence
of the three opinions). 
This mean field scenario essentially also holds
in a finite population when $\zeta\geq\Delta_c$, as illustrated by Fig.~\ref{Fig4}(a),
where at fixed $N$ it is found that
\begin{itemize}
 \item  the centrism consensus time  $\tau$ grows when the density  $z$ of susceptible centrists decays;
\item $\tau$ decreases monotonically when  $\delta$ decreases and $\zeta$ increases;
\item at fixed value of $\zeta-\delta$, $\tau$ increases monotonically when  $\delta$ and $\zeta$ decrease;
\item  the parameters $\zeta$ and $\delta$ appear to have only a weak effect on the MCT when $z\to 0$;
\item the predictions (\ref{sol1}) obtained from the backward Fokker-Planck equations  are in excellent agreement with the 
results of stochastic simulations. 
\end{itemize}
When $\zeta>\Delta_c$, the integrals in (\ref{sol1})  are difficult to evaluate analytically 
since the  function ${\rm exp}\left(-N {\cal F}(u)\right)$ has no isolated peaks and thus no saddle-point approximation 
can be carried out. Nevertheless (\ref{sol1}) can be studied numerically and it is found
that when $z\to 0$, i.e. when the initial population consists only of $A$ and $B$ voters (along with centrist zealots,
 but no susceptible 
centrists), one has
\begin{eqnarray}
 \tau \sim \ln{N}, \quad \text{when} \; N\gg 1.
\end{eqnarray}
This result, compatible with the above points (iv) and (v), is confirmed by 
 Fig.~\ref{Fig4}(b) where, when $N$ is large and increases, it is shown that
\begin{itemize}
 \item  $\tau(0)$ grows linearly with $\log_{\rm 10} N$, with a slope that appears to slightly
increase  when $\zeta$ and $\delta$ decrease, as shown in Fig.~\ref{Fig4}(b) when $N\geq 1000$;
\item the formula  (\ref{sol1}) fully captures the dependence of $\tau$ on $N$ for all
population sizes in the range $N=10^2-10^4$.
\end{itemize}
\section{Mean consensus time in the 3CVM with centrist zealots and asymmetric persuasion biases}
When the biases toward $A$ and $B$ voters are different, i.e. when $\delta_A>\delta_B>0$,
the system's dynamics is described by a bivariate birth-death process characterized by 
the master equation (\ref{MEbiv}).  The mean time to reach centrism consensus 
starting with a population comprising a fraction of 
$x$ voters of type $A$ and a density $y$ of $B$-voters thus
 obeys 
the following backward master equation~\cite{Gardiner}:
\begin{eqnarray}
 \label{backME2}
\tau(x,y)&=& \Delta +
T_x^{-}\tau(x-\Delta ,y) + T_x^{+}\tau(x+\Delta ,y) +
T_y^{-}\tau(x,y-\Delta ) + T_y^{+}\tau(x,y+\Delta )\nonumber\\ &+&
\left[1-\left(
T_x^{+}+T_x^{-}+T_y^{+}+T_y^{-}
\right)\right] \tau(x,y),
\end{eqnarray}
where $\Delta =N^{-1}$. Since we consider a  finite but large population size ($N\gg 1$), 
the transition rates  $T_{x,y}^{\pm}$
are the continuous versions of (\ref{TRbiv}), i.e. $T_{x}^{+}\equiv (1+\delta_A)x z$, $T_{x}^{-}\equiv x(z+\zeta)$,
$T_{y}^{+}\equiv (1+\delta_B)y z$ and $T_{y}^{-}\equiv y(z+\zeta)$.
 It is generally difficult to solve (\ref{backME2})
and even a size expansion is not very useful to make analytical progress.
Yet, the bivariate master equation (\ref{backME2}) can be simplified 
in the limits where $x\to 0$ and $y\to 0$, and
the asymptotic behavior of the MCT can be obtained when the population size is large
 (see below). In fact,
when $y=0$ and $x=1-\zeta-z>0$,  $T_{y}^{\pm}=0$ while
$T_{x}^{\pm}=T^{\mp}(z)$ and therefore $\tau(x,0)=\tau_{\delta_A}(z=1-\zeta-x)$, where $\tau_{\delta_A}$
is the solution of the single-variate master equation (\ref{backME}) with $\delta$ replaced by $\delta_A$.
Similarly, when $x=0$ and  $y=1-\zeta-z>0$, one has $\tau(0,y)=\tau_{\delta_B}(z=1-\zeta-y)$.
Building on our knowledge of the case $\delta_A=\delta_B$ (Sec.~6), we know that the solution  of (\ref{backME}) increases 
monotonically with $\delta$ when $\zeta$ is kept fixed (see Fig.~\ref{Fig2}). Since $\delta_A>\delta_B$, this implies 
$\tau(x,0)=\tau_{\delta_A}(z=1-\zeta-x)>\tau(0,y)=\tau_{\delta_B}(z=1-\zeta-y)$. 
\\
More generally,
at fixed values of $z$ and $\zeta$, the largest MCT is attained when the population of non-centrists consists
of a maximum  number of most persuasive radical voters, i.e.
 $N-\ell-\ell_{\zeta}$ voters of species  $A$.
Similarly, the shortest MCT at fixed values of $z$ and $\zeta$ 
is obtained when the initial sub-population of non-centrists  only consists of $N-\ell-\ell_{\zeta}$ voters of species $B$.
Hence, the MCT respectively increases and decreases with the initial fraction of $A$-voters
and $B$-voters in the population,  and 
the quantities $\tau_{\delta_A}$ and $\tau_{\delta_B}$
are upper and lower bounds of the MCT [see Eqs~(\ref{taudelta},\ref{Fdelta})]:
\begin{eqnarray}
\label{tauxy}
\tau_{\delta_B}(1-\zeta-x-y)\leq\tau(x,y)\leq \tau_{\delta_A}(1-\zeta-x-y),
\end{eqnarray}
as illustrated by Figs.~\ref{Fig5} and \ref{Fig6}, see below.
According to this discussion, we notice that a quantitative difference between the cases of asymmetric and identical 
persuasion biases lies in the influence of the 
 initial condition: when $\delta_A\neq \delta_B$, the MCT depends on the specific initial densities  $x$ of $A$'s and $y$
of $B$'s (not only on their sum, as in the symmetric case with $\delta_A=\delta_B$). 
In what follows, we show that when $z\ll 1-\zeta$ the asymptotic behavior
 of the MCT is independent 
of the initial condition when $N(\delta_A-\delta_B)
\gg 1$ and $N(\delta_A -\zeta)\gg 1$: In this situation significant insight into its properties can be gained from the results of Sec.~6 combined with the
mean field analysis of Sec.~4.
\subsection{Long-lived coexistence and consensus time when $\zeta<\Delta_c$ and $\delta_A>\delta_B$}
\begin{figure}
\begin{center}
\includegraphics[width=2.2in, height=1.665in,clip=]{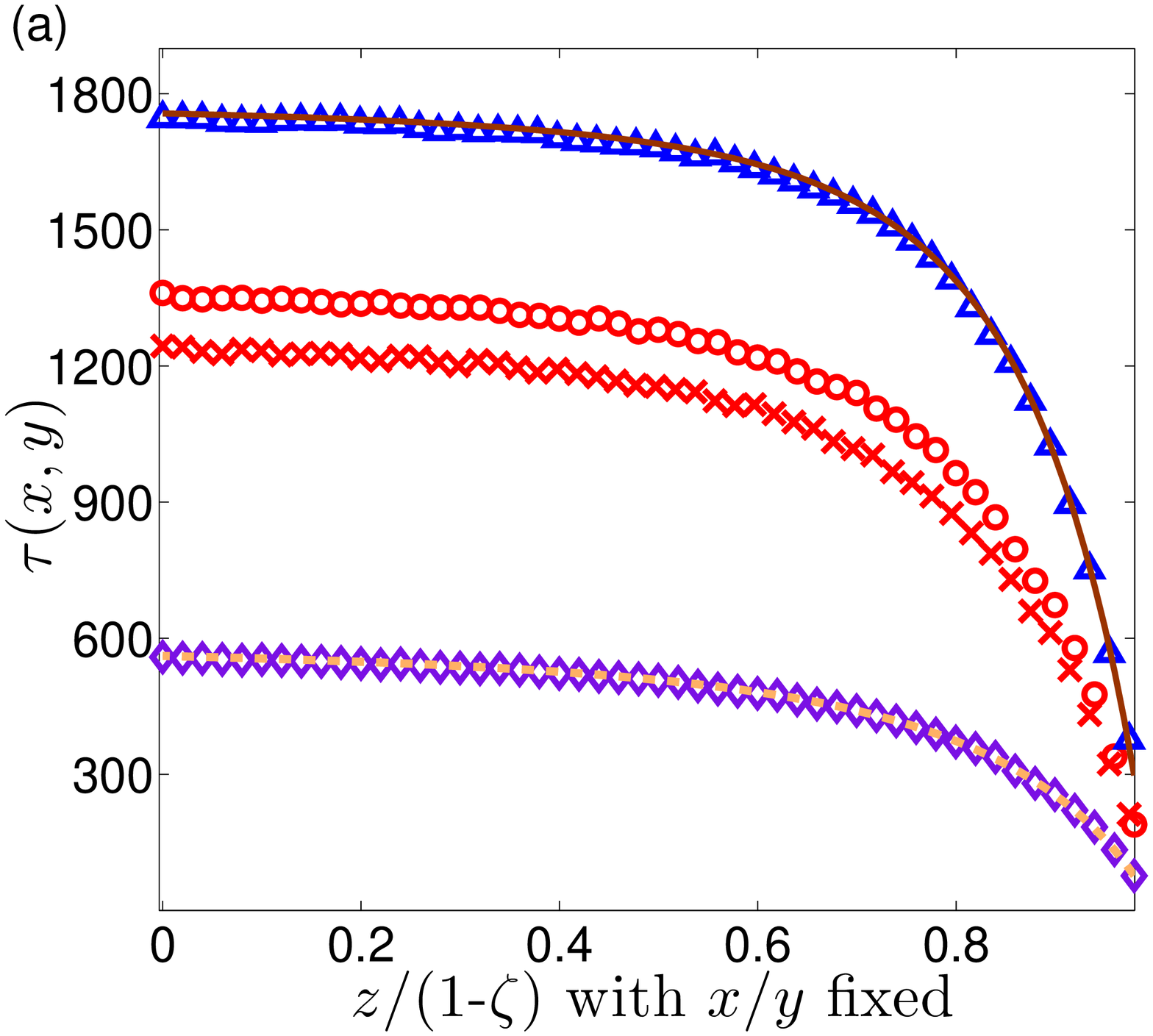}
\includegraphics[width=2.2in, height=1.7in,clip=]{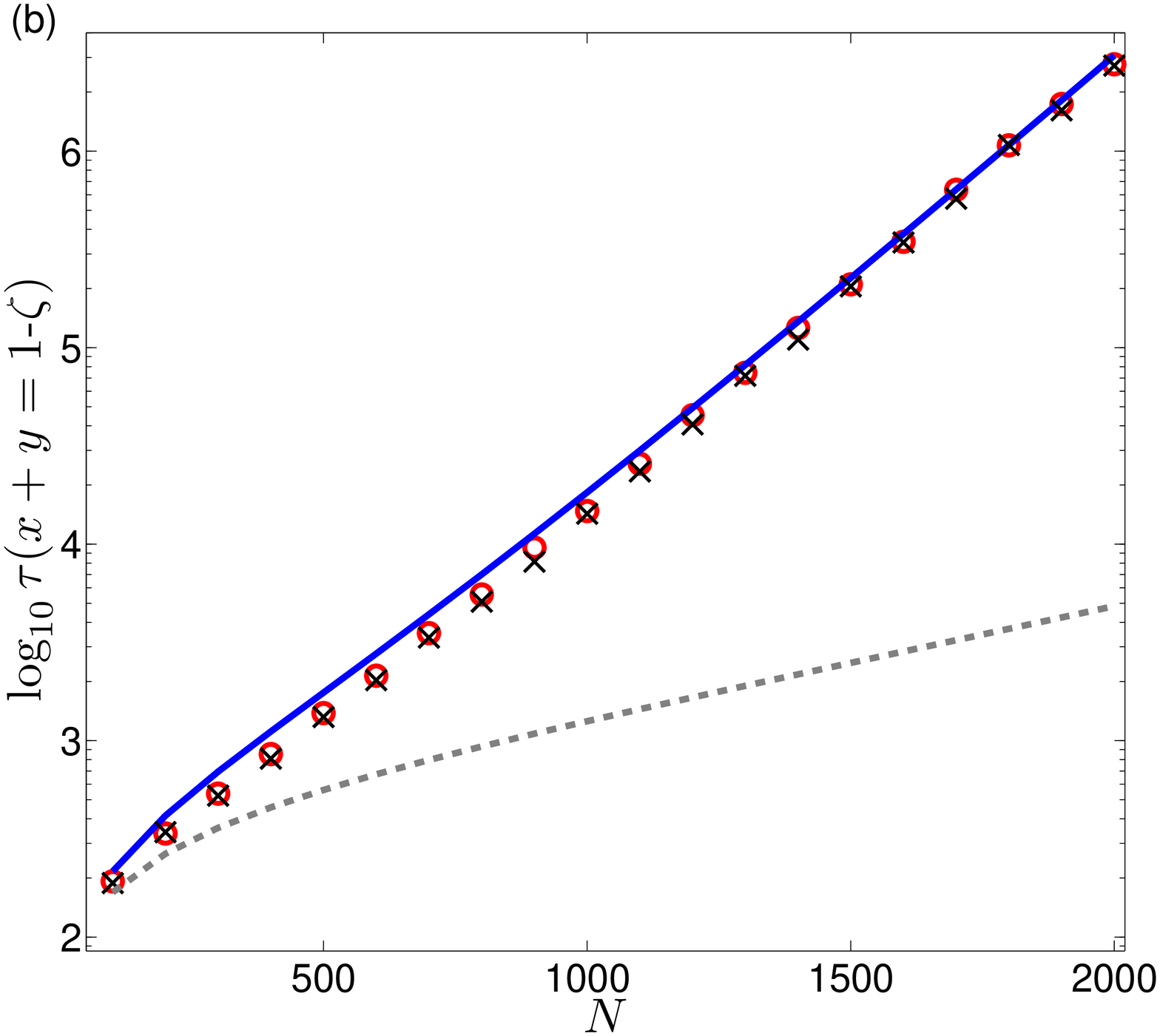}
\caption{{\it (Color online)}. Functional dependence of the mean consensus time when 
$\zeta<\Delta_c$ and $\delta_A>\delta_B$, with $\zeta=0.02$.
(a) The MCT as function of the rescaled initial density  $z/(1-\zeta)$ with the ratio $x/y$ held fixed, see text. 
Here, $(\delta_A,\delta_B)=(0.04,0.03)$   with $x=y~(\circ)$ and $y=1.5x~(\times)$, the population size is $N=500$. 
The results of stochastic simulations (symbols) 
averaged over $2.5\times 10^4$ samples
are compared with the results obtained for $\delta_A=\delta_B=0.04$ ($\triangle$, solid)
and  $\delta_A=\delta_B=0.03$ ($\diamond$, dashed). The analytical results (solid and dashed lines) are based on (\ref{taudelta},\ref{Fdelta}). 
(b) Logarithm of the MCT as  function of  $N$ (with $N=100-2000$) for the same parameters as in (a): $\log_{10}{\tau(x+y=1-\zeta)}$
 computed using stochastic simulations ($\circ$,~$\times$) averaged over $10^3$ samples 
is compared with the predictions of
(\ref{taudelta},\ref{Fdelta}) for $\delta=0.04$ (solid)
and $\delta=0.03$ (dashed).  
}
  \label{Fig5}
\end{center}
\end{figure} 
When $\zeta<\Delta_c\equiv \delta_A/(1+\delta_A)$, the mean field analysis (Sec.~4) has revealed the asymptotic stability of the state
where $A$-voters and centrists coexist.
As seen in Sec.~6.1, when the population size is large but finite, we expect the coexistence state to be 
long-lived and to decay after a time growing exponentially with the population size $N$ (to leading order).
If the population initial composition is $(x,y,z)=(1-\zeta,0,0)$ and  only consists of $A$-voters and
centrist zealots, one recovers a single-variate process (as in Sec.~6) and the MCT is 
$\tau(x=1-\zeta,0)=\tau_{\delta_A}(z=0)$,
where, in the realm of the Fokker-Planck equation, $\tau_{\delta_A}(z)$ is directly inferred from (\ref{sol1})
using
\begin{eqnarray}
\label{taudelta}
\tau_{\delta}(z)=2N \, \int_{z}^{1-\zeta} dv \; e^{N{\cal F}_{\delta}(v)}
\int_{0}^{v} du \, \frac{e^{-N{\cal F}_{\delta}(u)}}{\left( 1-\zeta-u\right)\left(\zeta+(2+\delta)u\right)},
\end{eqnarray}
with
\begin{eqnarray}
\label{Fdelta}
{\cal F}_{\delta}(u)&\equiv& \left(\frac{2\delta}{2+\delta}\right)u-
\frac{4\zeta(1+\delta)}{(2+\delta)^2}\ln{\left[(2+\delta)u + \zeta\right]}.
\end{eqnarray}
Similarly, if the initial state is 
$(x,y,z)=(0,1-\zeta,0)$, one has $\tau(0,1-\zeta)=\tau_{\delta_B}(0)$.
\\
 Since 
 $\delta_A>\delta_B$ and $\tau_{\delta}\sim N^{-1/2} e^{N\gamma(\delta,\zeta)}$, with an exponent
$\gamma={\cal F}_{\delta}(1-\zeta)-{\cal F}_{\delta}(\zeta/\delta)$ that grows monotonically with $\delta$ 
when $\zeta$ is fixed (see Sec.~6.1 and Fig.~\ref{gammafig}), one has $\tau(x,0)=\tau_{\delta_A}(0)\sim N^{-1/2} e^{N\gamma(\delta_A,\zeta)}\gg \tau(0,y)=
\tau_{\delta_B}(0)$ when $N(\delta_A-\delta_B)\gg 1$,
$N(\delta_A -\zeta)\gg 1$, and the initial density  of susceptible centrists is vanishingly small ($z\to 0$). 

In the  case of a population with an arbitrary initial density $(x,y,z)$ of  $A,B$ and $C$'s, 
the system almost surely reaches the metastable 
state $(a_2^*,b_2^*,c_2^*)=(1-\zeta-\zeta/\delta_A,0,\zeta/\delta_A)$, as prescribed by the mean 
field equations (\ref{RE}) [see (\ref{fixed1}) and Fig.~\ref{MFdiag1}(a)], before reaching  centrism consensus. 
Hence, the less persuasive of the radical voters ($B$'s)
quickly disappear from the population and the
 metastable state only consists of centrists and
voters of the most persuasive party ($A$'s).
Apart from the short transient necessary to reach the metastable state,  the  
MCT is thus obtained from
the mean time to reach consensus starting from the fixed point $(a_2^*,b_2^*,c_2^*)$
where the population comprises only $A$-voters and centrists. 
In this situation, 
when $N(\delta_A-\delta_B)\gg 1$ and $N(\delta_A -\zeta)\gg 1$, the leading contribution to the
mean consensus time $\tau(x,y)$ is 
obtained from a single-variate backward master equation [like (\ref{backME}) of Sec.~6] 
and is given by $\tau_{\delta_A}(z)$ with (\ref{taudelta}).
 In particular, when the initial 
density $z$ of susceptible centrists 
is very small (i.e. $z\ll 1-\zeta$) with $N(\delta_A-\delta_B)\gg 1$ and $N(\delta_A -\zeta)\gg 1$, 
one has
$\tau(x+y\simeq 1-\zeta)
\simeq \tau_{\delta_A}(0)$ and, from the results
 (\ref{sol4}) and (\ref{MET1}), we infer the leading contribution to the MCT
when $\delta_B<\delta_A\ll 1$ and $\zeta\ll 1$:
\begin{eqnarray}
\label{tauxyas}
\tau(x+y\simeq 1-\zeta)\simeq \tau_{\delta_A}(0)\sim N^{-1/2}\,e^{N(\delta_A- \zeta)}\left(\frac{\zeta}{\delta_A}\right)^{N\zeta}.
\end{eqnarray}
The results of stochastic simulations  reported in Fig.~\ref{Fig5} corroborate the above analysis. In fact, Fig.~\ref{Fig5}(a) reveals the dependence of  
the MCT on the initial conditions: $\tau$ grows monotonically with $z^{-1}$
and  $x/y$, but exhibits only a weak dependence on $x/y$. Moreover, the comparison 
with the case of identical biases illustrates that  $\tau_{\delta_B}(z=1-\zeta-x-y)\leq \tau(x,y)\leq \tau_{\delta_A}(z=1-\zeta-x-y)$
which
confirms (\ref{tauxy}).
Fig.~\ref{Fig5}(b) shows that asymptotically 
the MCT grows exponentially  (to leading order) with the population size  when $N(\delta_A-\delta_B)\gg 1$,
$N(\delta_A -\zeta)\gg 1$
and $z\to 0$, 
in full agreement with the behavior $\tau(x+y=1-\zeta)\simeq \tau_{\delta_A}(0)$ 
predicted by
(\ref{taudelta})-(\ref{tauxyas}). The
figure~\ref{Fig5}(b) also illustrates that the asymptotic behavior  of the MCT
is independent of the ratio $x/y$ and therefore of the detailed composition of the initial population.
\subsection{Mean consensus time when $\zeta\geq \Delta_c$ and $\delta_A>\delta_B$}
\begin{figure}
\begin{center}
\includegraphics[width=2.19in, height=1.69in,clip=]{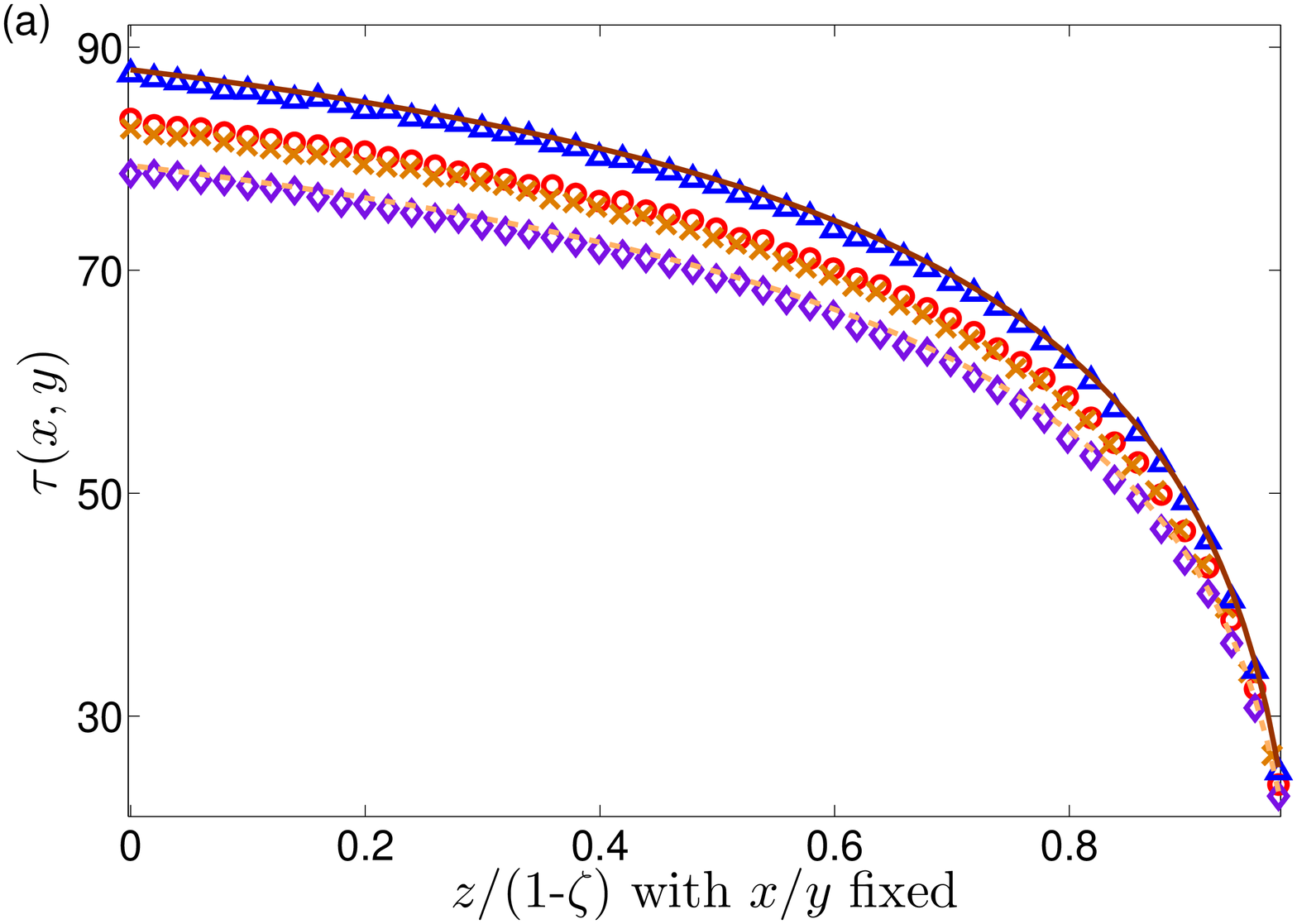}
\includegraphics[width=2.19in, height=1.69in,clip=]{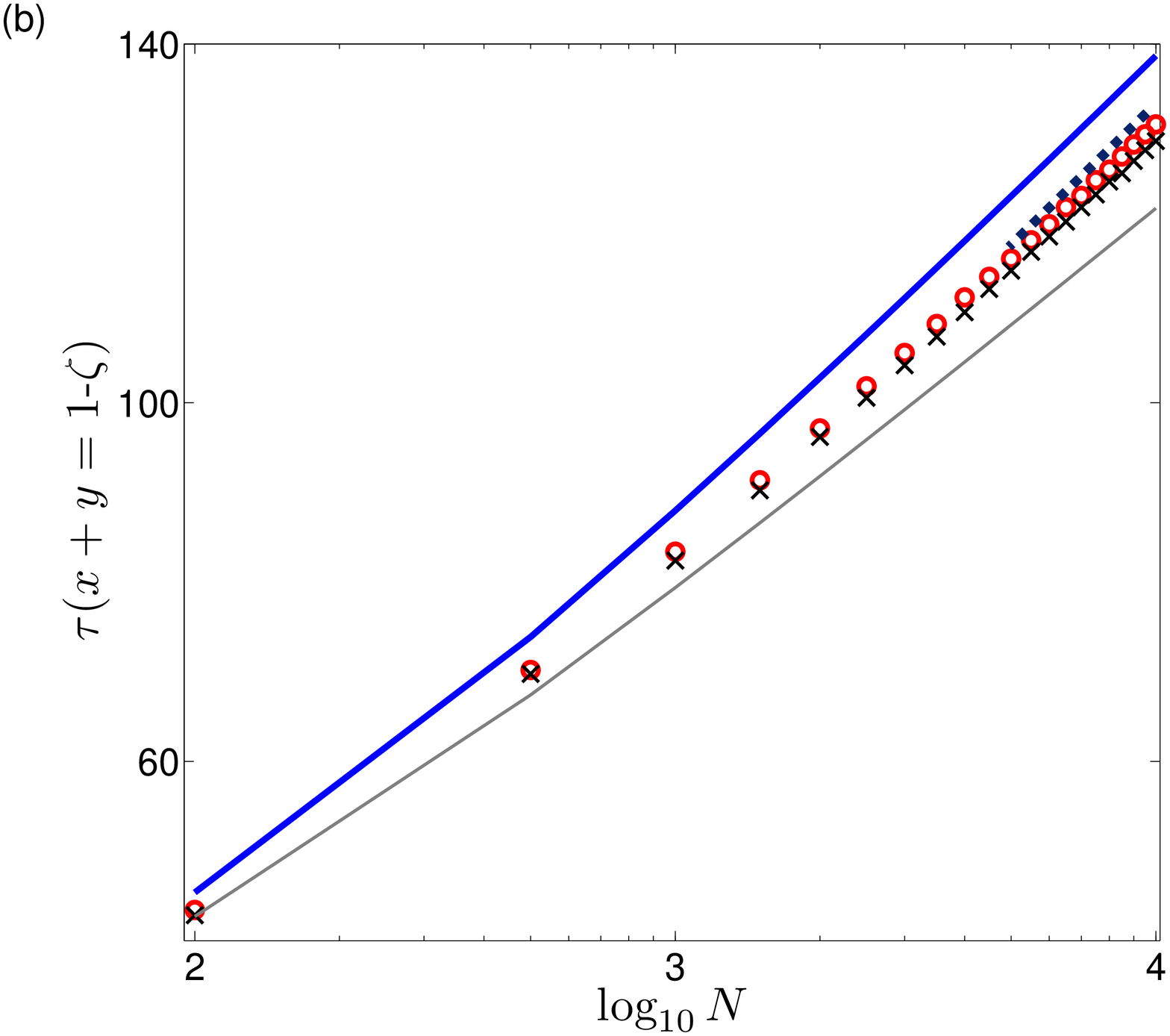}
\caption{{\it (Color online)}. Functional dependence of the MCT when $\zeta\geq\Delta_c$ and
 $\delta_A>\delta_B$, with $\zeta=0.08$.
(a) The MCT as function of the rescaled initial density  $z/(1-\zeta)$ with the ratio $x/y$ held fixed, see text. Here, $(\delta_A,\delta_B)=(0.04,0.03)$   with $x=y~(\circ)$ and $y=1.5x~(\times)$, the population size is $N=1000$. 
The results of stochastic simulations (symbols), 
averaged over $3.5\times 10^4$ samples,
are compared with results obtained for $\delta_A=\delta_B=\delta=0.04$ ($\triangle$,  solid)
and  $\delta=0.03$ ($\diamond$, dashed).
The analytical results (solid and dashed) are based on (\ref{taudelta},\ref{Fdelta}). 
(b) $\tau(x+y=1-\zeta)$ as  function of  $N$ in a semi-logarithmic scale for the same parameters as in (a).
Results of stochastic simulations (symbols) for $N=100-10000$
are compared with
$\tau_{\delta_B}(0)$ (thin solid)
and $\tau_{\delta_A}(0)$ (thick and, as a guide for the eyes, dashed line)
obtained from (\ref{taudelta},\ref{Fdelta}), see text.   
}
  \label{Fig6}
\end{center}
\end{figure} 
When  $\zeta \geq \Delta_c$, the only stable fixed point corresponds to centrism consensus.
From the analysis of Sec.~6.2, we know that in a finite  population 
with initial composition $(x=1-\zeta,0,0)$, the MCT is $\tau(1-\zeta,0)= \tau_{\delta_A}(z=0)$ and 
can be obtained from (\ref{taudelta}). Similarly,  
for an initial composition  $(0,y,0)$, the  MCT is $\tau(0,1-\zeta)= \tau_{\delta_B}(z= 0)$.
When $N\gg 1$ and the initial number of susceptible centrists is small, i.e. $z=1-\zeta-x-y \to 0$, we have seen in Sec.~5.2
(see Fig.~\ref{Fig4}(b)) that $\tau_{\delta_A}(0) \sim \tau_{\delta_B}(0) \sim \ln{N}$. Furthermore, according to  (\ref{tauxy}) 
 $\tau_{\delta_A}$ and $\tau_{\delta_B}$
are upper and lower bounds of the MCT. The results of Sec.~6.2 hence imply 
\begin{eqnarray}
\label{tauxybis}
&&\tau_{\delta_B}(0) \leq \tau(x+y\simeq 1-\zeta)\leq \tau_{\delta_A}(0) , \quad \text{and therefore} \nonumber\\
&&\tau(x+y\simeq 1-\zeta) \sim \tau_{\delta_A}(0)\sim \tau_{\delta_B}(0) \sim \ln{N}, \quad \text{when} \;  N\gg 1.
\end{eqnarray}

These findings are corroborated by the  results of stochastic simulations  reported in Fig.~\ref{Fig6} 
that confirms (\ref{tauxy}):  $\tau_{\delta_B}(z=1-\zeta-x-y)\leq \tau(x,y)\leq \tau_{\delta_A}(z=1-\zeta-x-y)$. 
The dependence of the MCT on the initial conditions is illustrated in 
Fig.~\ref{Fig6}(a) where  $\tau(x,y)$ grows monotonically with  $z^{-1}$
and  $x/y$ (held constant). We notice the dependence on $x/y$ is 
particularly weak.
 Fig.~\ref{Fig6}(b) confirms the logarithmic growth with $N$ of  
 the MCT  when the population size becomes very large,
in full agreement (\ref{tauxybis}). In particular, the symbols in Fig.~\ref{Fig6}(b)
are found to be aligned parallel to the thick and dashed lines when $N\gg 1$,
which implies  that
asymptotically $e^{\tau(x,y)}\simeq Q(x,y) e^{\tau_{\delta_A}(0)}$ or $\tau(x,y)\simeq \tau_{\delta_A}(0)+ 
\ln{Q(x,y)}  $, where the function $Q$  
depends weakly on the initial densities $(x,y)$. 

\section{Conclusion and discussion}
After having reviewed  the main properties of the paradigmatic voter model and of some of its generalizations accounting for zealotry, 
commitment and incompatibility (Section~2), we have studied how a committed minority may resist a persuasive majority and 
how the resulting competition  influences the maintenance of cultural diversity.
Here, this line of investigation has been carried out in the framework of the three-party constrained voter model
with ``centrist zealots''.
 In fact, following the basic idea of the bounded-compromise and three-party constrained 
voter  models, we have considered that individuals of two radical parties ($A$ and $B$) are incompatible among them and 
compete to impose their consensus by convincing centrists that, in turn, can convince $A$'s and $B$'s.
To reflect the fact that radical ($A$ and $B$) voters are more persuasive than centrists, 
we have considered that there are {\it persuasion biases} $\delta_A$ and $\delta_B$ toward opinions $A$ and $B$
(with $\delta_A\geq \delta_B$). Furthermore, we  have also  assumed that the population includes 
susceptible centrists ($C$) and a small fraction $\zeta$ of {\it centrist zealots} ($C_{\zeta}$).
Whereas $C$'s can radicalize and adopt either the state $A$ or $B$, centrist zealots are committed individuals that always 
remain centrists and oppose the formation of any consensus that is not centrism.
Hence, the  persuasion of  $A$'s and $B$'s is resisted and opposed by
the commitment of centrist zealots. This results in a subtle competition between commitment and persuasion that has 
been studied in the mean field limit and in a well-mixed  population  of finite size $N$ (on a complete graph).
 
In an infinitely large population, there is a continuous transition between a coexistence  (``multicultural'') phase, that is stable
 when the fraction of centrist zealots is below the critical threshold $\Delta_c=\delta_A/(1+\delta_A)$,
and a phase where centrism prevails when $\zeta\geq \Delta_c$. This scenario changes in finite populations 
 when demographic fluctuations ultimately lead to  centrism consensus. The competition between commitment and persuasion is thus 
characterized by the mean consensus time and has been investigated in terms of  single-variate and bivariate birth-death processes.
The case of identical persuasion bias has been thoroughly analyzed and the mean consensus time has been computed in the realm of the 
diffusion theory (Fokker-Planck equation) and with the WKB method. In a large population, it has been shown that the 
long-time dynamics of the case with  asymmetric biases 
($\delta_A>\delta_B>0$) can be described in terms of a single-variate birth-death process
involving only centrists and $A$-voters. Hence, the asymptotic behavior of the mean 
consensus time in a large population  and its upper and lower bounds have been obtained analytically also when $\delta_A>\delta_B$. 
In fact, the main difference between the cases $\delta_A=\delta_B$ and $\delta_A>\delta_B$ lies in the composition of 
the metastable state: 
it is characterized by the coexistence of   $A$-voters and centrists (no $B$-voters)
 when  $\delta_A>\delta_B$, whereas voters of the three parties coexist when the persuasion biases are identical 
 ($\delta_A=\delta_B$).

 Our findings,  can  be summarized as follows:
When the fraction of centrist zealots in the population of size $N$ is low ($\zeta<\Delta_c$), we have shown that 
the interplay between commitment and persuasion results in a metastable coexistence (multicultural) phase. 
The mean consensus time thus corresponds to the decay rate of the metastable state 
and, when centrists are initially in the minority, it has been shown to grow as $\tau\sim N^{-1/2}e^{N\gamma(\delta_A,\zeta)}$ with an 
exponent $\gamma$ determined 
explicitly. When the biases are asymmetric, the less persuasive of the radical voters ($B$'s)
quickly disappear from the population and the
 metastable state only consists of centrists and
voters of the most persuasive party ($A$'s), whereas opinions of all parties coexist when $\delta_A=\delta_B$.
On the other hand, when  $\zeta\geq\Delta_c$, the system attains centrism consensus in a time that scales logarithmically with the population size, i.e. $\tau\sim \ln{N}$.
These features are found to be robust since they neither depend of the detailed initial composition of the population nor on the 
weakest persuasion bias.
It is  worth noticing that a similar type of behavior 
concerning the mean fixation time has recently been found in evolutionary games and in models of population 
genetics, see e.g.~Refs.~\cite{AM2011,Games}.

In conclusion, while the mean consensus time of the three-party constrained voter model without zealots scales linearly with $N$,  
we have here shown that the presence of small fraction of
zealots and the existence of persuasion biases results
in a rich dynamics characterized either by a prolonged maintenance of a multicultural phase, or   
by a quick realization of centrism consensus. 
The long-lived coexistence state generally consists of voters of the party supported by the most committed individuals
along with the most persuasive voters. Since the description of the (meta-)stability of the multicultural phase and the mean 
time to reach the consensus  
cannot be obtained solely from the mean field rate equations, this work illustrates the relevance of statistical physics methods to study problems of ``opinion dynamics''.

\begin{acknowledgements}
The author is grateful to Sid Redner for
 useful discussions and insightful comments on a preliminary version of the manuscript.
\end{acknowledgements}


\begin{thebibliography}{}
%
%
%
\bibitem{Axelrod}
 R. Axelrod, J. Conflict Resolution 41, 203 (1997); R. Axelrod, {\it The complexity of cooperation}, (Princeton University
Press, 1997);
 C. Castellano, M. Marsili, and A. Vespignani, Phys. Rev. Lett. {\bf 85}, 3536 (2000); 
K. Klemm, V.~M. Eguiluz, R. Toral, and M. San Miguel, Phys.~Rev.~E {\bf 67}, 045101 (2003).
%
\bibitem{BoundedCompromise}
G.~Deffuant, D.~Neau, F.~Amblard, G.~Weisbuch, Adv. Complex Syst. {\bf 3}, 87 (2000); 
R.~Hegselmann and U. Krause, Journal of Artificial Societies and Social Simulation {\bf 5}, issue 3 (2002);
G.~Weisbuch, G~Deffuant, F.~Amblard, and J.-P.~Nadal, Complexity {\bf 7}, 55 (2002);
E. Ben-Naim, P.~L. Krapivsky, 
and S. Redner, Physica D {\bf 183}, 190 (2003);
F. Slanina,  Eur. Phys. J. B {\bf 79}, 99 (2011).
%
%
\bibitem{SocRev}
C. Castellano, S. Fortunato, and V. Loreto, Rev. Mod. Phys. {\bf 81}, 591 (2009).
%
\bibitem{PopGen} J.~F. Crow and M. Kimura, {\it An Introduction to Population Genetics Theory} (The Blackburn Press, New Jersey, 1970);
W.~J. Ewens, {\it mathematical Population Genetics} (Springer, U.S.A., 2004);
 M. A. Nowak, {\it Evolutionary Dynamics} (Belknap Press, 2006); 
R.~A. Blythe and A.~J. McKane, J.~Stat.~Mech. {\bf P07018} (2007);
G. Szab\'o  and G. F\'ath, Phys. Rep. {\bf 446} 97 (2007).
%
%
\bibitem{voter-variants}
S. Galam, J.~Stat.~Phys. {\bf 61}, 943 (1990); 
Physica A {\bf 274}, 132 (1999); Eur.~Phys.~J. B {\bf 25}, 403 (2002);
Eur. Phys. J. B {\bf 45}, 569 (2005);
K. Sznajd-Weron, J. Sznajd, Int.~J. Mod.~Phys. C {\bf 11}, 1157
(2000);
M. Mobilia and S. Redner,
Phys.~Rev.~E {\bf 68}, 046106 (2003); 
%
R. Lambiotte and S. Redner, EPL {\bf 82}, 18007 (2008);
F.~Slanina, K.~Sznajd-Weron, and D.~Przybyla,
EPL {\bf 82} (2008) 18006;
I.~J. Benczik, S.~Z. Benczik, B. Schmittmann, and R. K. P. Zia, EPL {\bf 82}, 48006 (2008); 
D. Volovik, M. Mobilia, and S. Redner, EPL {\bf 85},  48003 (2009).
%
%
\bibitem{Liggett}
T. M. Liggett, {\it Interacting Particle Systems} (Springer, U.S.A., 1985).
%
%
\bibitem{Schelling}
T.~C. Schelling, J.~Math.~Social.~{\bf 1}, 143 (1971);
{\it Micromotives and Macrobehavior} (Norton, New York, 1978).
%
%
\bibitem{Granovetter}
M.~Granovetter, Am.~J.~Sociol. {\bf 78}, 1360 (1973); {\it ibid.} {\bf 83}, 1420 (1978).
%
%
\bibitem{VKR03}
F. Vazquez, P.~L. Krapivsky,  and S. Redner, J.~Phys.~A:~Math.~Gen. {\bf 36}, L61 (2003).
\bibitem{Redner04}
F. Vazquez and S. Redner, J.~Phys.A:~Math.~Gen. {\bf 37}, 8479 (2004).
%
\bibitem{MM11}
M. Mobilia, EPL {\bf 95},50002 (2011).
%
\bibitem{Lanchier12}
N. Lanchier, Ann.~Appl.~Prob. {\bf 22}, 860 (2012). 
%
\bibitem{zealots1}
M. Mobilia, Phys.~Rev.~Lett. {\bf 91},  028701 (2003).
\bibitem{zealots2}
M. Mobilia and I.~T. Georgiev,
Phys.~Rev. E {\bf 71}, 046102 (2005).
\bibitem{zealots3}
M. Mobilia, A. Petersen, and S. Redner, J. Stat. Mech., {\bf P08029} (2007).
%
\bibitem{GalamJacobs}
S. Galam and F. Jacobs, Physica A {\bf 381}, 366 (2007).
%
%
\bibitem{committed}
J.~Xie, S.~Sreenivasan, G.~Korniss, W.~Zhang, C.~Lim, and B.~K.~Szymanski,
Phys. Rev. E {\bf 84}, 011130 (2011).
%
%
\bibitem{Glauber}
R.~J. Glauber, J.~Math.~Phys. {\bf 4}, 294 (1963).
%
\bibitem{KRB}
P.~L. Krapivsky, S.~Redner and E.~Ben-Naim, {\it A kinetic view of statistical physics} (Cambridge, New York, 2010).
%
%
\bibitem{Moran}
P.~A.~P. Moran, {\it The statistical processes of evolutionary theory} (Clarendon, Oxford, 1962).
%
%
\bibitem{VMnets}
V. Sood and S. Redner, Phys.~Rev.~Lett. {\bf 94}, 178701 (2005);
T. Antal, V. Sood, and S. Redner, Phys.~Rev.~Lett. {\bf 96}, 188104 (2006);
V. Sood, T. Antal, and S. Redner, Phys.~Rev.~E {\bf 77}, 041121 (2008);
G. J. Baxter, R. A. Blythe, and A. J. McKane,
Phys. Rev. Lett. 101, 258701 (2008);
R.~A. Blythe, J. Phys A: Math. Theor. {\bf 43} 385003 (2010).
%
%
\bibitem{Gardiner} C. W. Gardiner, {\it Handbook of Stochastic Methods} (Springer,  U.S.A., 2002); N. G. van Kampen, 
{\it Stochastic Processes in Physics and Chemistry}, (North-Holland, Amsterdam, 1992); S. Redner, {\it A Guide to First-Passage Processes} (Cambridge University Press, U.S.A., 2001);
H. Risken, {\it The Fokker-Planck equation},  (Springer, New York, 1989);
S. Karlin and H.~M. Taylor, {\it A Second Course in Stochastic Processes} (Academic Press, New York, 1981).
%
\bibitem{AM2012}
M. Assaf and M. Mobilia, Phys.~Rev.~Lett. {\bf 109}, 188701 (2012).
%
\bibitem{Coarsening}
P.~L. Krapivsky, Phys.~Rev. A {\bf 45}, 1067 (1992); E.~Ben-Naim, L.~Frachebourg, and P.~L. Krapivsky,
Phys. Rev. E {\bf 53}, 3078 (1996); L.~Frachebourg and P.~L. Krapivsky, Phys. Rev. E {\bf 53}, R3009 (1996).
%
\bibitem{Commit}
D. Volovik and S. Redner, J.~Stat.~Mech. {\bf P04003} (2012).
%
\bibitem{Heterogeneity}
N. Masuda, N. Gibert, and S. Redner, Phys.~Rev.~E {\bf 82}, 010103 (2010)
%
\bibitem{EGT-heterogeneous}
M. Mobilia, Phys.~Rev. E 86, 011134 (2012);
N. Masuda, Sci.~Rep. {\bf 2}, 646 (2012).
%
\bibitem{NG}
L.~Steels, Artif. Life {\bf 2}, 319 (2005);
A.~Baronchelli,  M.~Felici, E.~Caglioti, V.~Loreto, and L~Steels, J.~Stat.~Mech. (2006) {\bf P06014}. 
%
%
\bibitem{Kimura} J.~F. Crow and M. Kimura, {\it An Introduction to Population Genetics Theory} (Blackburn Press, New Jersey, 2009);
W.~J. Ewens, {\it Mathematical Population Genetics} (Springer, New York, 2004).
%
%
\bibitem{Gillespie}
D.~T. Gillespie, J. Comput. Phys. {\bf 22}, 403 (1976).

\bibitem{Assaf2010}
M.~Assaf and B.~Meerson, 
Phys. Rev. Lett., {\bf 97}, 200602  (2006);
Phys. Rev. E, {\bf 75}, 031122 (2007);
Phys. Rev. E {\bf 81}, 021116 (2010).
%
\bibitem{AM2011}
M.~Mobilia and M.~Assaf, EPL {\bf 91}, 10002 (2010);
M.~Assaf and M.~Mobilia, J. Stat. Mech. {\bf P09009} (2010);
M.~Assaf and M.~Mobilia, J. Theor. Biol. {\bf 275}, 93 (2011).
%

\bibitem{WKB}
L.~D. Landau and E.~M. Lifshitz, ``Quantum Mechanics:~Non-Relativistic Theory''
(Pergamon, London, 1977).
%
\bibitem{Dykman}
R.~Kubo, K.~Matsuo, and K.~Kitahara,
J.~Stat.~Phys. {9}, 51 (1973);
M.~I. Dykman, E.~Mori, J.~Ross,  P.~M. Hunt, J. Chem. Phys. {\bf 100}, 5735 (1973).
%
\bibitem{Escudero}
C. Escudero and A. Kamenev,
Phys. Rev. E {\bf 79}, 041149 (2009).
%
%
\bibitem{Games}
T.~Antal and I.~Scheuring, Bul.~Math.~Biol. {\bf 68}, 1923 (2006); 
J.~ Cremer, T.~Reichenbach, and E.~Frey, New J.~Phys. {\bf 11} 093029 (2009); 
M.~Mobilia, J.~Theor.~Biol. {\bf 264}, 1 (2010).
%
\end{thebibliography}


\end{document}